\documentclass[ ]{article}
\usepackage{amssymb}
\usepackage{amsmath,amssymb}
\usepackage{graphicx,epstopdf}
\usepackage{color}
\usepackage{subfigure}

\usepackage{hyperref}
\usepackage{enumitem}
\usepackage{xcolor}
\usepackage{xurl}
\usepackage{cite}
\usepackage[english]{babel}
\usepackage[normalem]{ulem}
\usepackage{booktabs}
\usepackage{float}
\usepackage{comment}
\usepackage{chngcntr}
\usepackage{rotating}
\usepackage{soul}
\usepackage{multirow}
\usepackage[font=it]{caption}
\usepackage{xurl} % allow line breaks anywhere in URL string

\definecolor{redcolor}{rgb}{1,0,0}

\definecolor{pinkcolor}{rgb}{0.85, 0.45, 0.6}

\definecolor{bluecolor}{rgb}{0,0,1}

\definecolor{blackcolor}{rgb}{0,0,0}
\def\Black#1{{\color{blackcolor} #1}}

\begin{document}

%\begin{frontmatter}

%% Title, authors and addresses

%% use the tnoteref command within \title for footnotes;
%% use the tnotetext command for the associated footnote;
%% use the fnref command within \author or \address for footnotes;
%% use the fntext command for the associated footnote;
%% use the corref command within \author for corresponding author footnotes;
%% use the cortext command for the associated footnote;
%% use the ead command for the email address,
%% and the form \ead[url] for the home page:
%%
%% \title{Title\tnoteref{label1}}
%% \tnotetext[label1]{}
%% \author{Name\corref{cor1}\fnref{label2}}
%% \ead{email address}
%% \ead[url]{home page}
%% \fntext[label2]{}
%% \cortext[cor1]{}
%% \address{Address\fnref{label3}}
%% \fntext[label3]{}

%\dochead{}
%% Use \dochead if there is an article header, e.g. \dochead{Short communication}
%% \dochead can also be used to include a conference title, if directed by the editors
%% e.g. \dochead{17th International Conference on Dynamical Processes in Excited States of Solids}

\title{Stylized Facts of High-Frequency Bitcoin Time Series}

%% use optional labels to link authors explicitly to addresses:
%% \author[label1,label2]{<author name>}
%% \address[label1]{<address>}
%% \address[label2]{<address>}
\author{
Yaoyue Tang$^{1}$\footnote{yaoyue.tang@sydney.edu.au},
Karina Arias-Calluari$^{2}$,
M. N. Najafi$^{3}$,\\
Michael S. Harr\'e$^{1}$ and
Fernando Alonso-Marroquin$^{4}$\footnote{fernando@quantumfi.net}
}
\date{}
\maketitle
%%%%%%%%% Insert author address here
\begin{center}
$^{1}$Modelling and Simulation Research Group, The University of Sydney, Sydney NSW 2006, Australia\\
$^{2}$School of Mathematics and Statistics, The University of Sydney, Sydney, NSW 2006, Australia\\
$^{3}$Department of Physics, University of Mohaghegh Ardabili, P.O. Box 179, Ardabil, Iran\\
$^{4}$CPG, King Fahd University of Petroleum and Minerals, Dhahran 31261,
Kingdom of Saudi Arabia
\end{center}
%%%% Subject entries to be placed here %%%%

\begin{abstract}

This paper analyses the high-frequency intraday Bitcoin dataset from 2019 to 2022. During this time frame, the Bitcoin market index exhibited two distinct periods, 2019-20 and 2021-22, characterized by an abrupt change in volatility. The Bitcoin price returns for both periods can be described by an anomalous diffusion process, transitioning from subdiffusion for short intervals to weak superdiffusion over longer time intervals. The characteristic features related to this anomalous behavior studied in the present paper include heavy tails, which can be described using a $q$-Gaussian distribution and correlations. When we sample the autocorrelation of absolute returns, we observe a power-law relationship, indicating time dependence in both periods initially. 

The ensemble autocorrelation of the returns decays rapidly. We fitted the autocorrelation with a power law to capture the decay and found that the second period experienced a slightly higher decay rate. The further study involves the analysis of endogenous effects within the Bitcoin time series, which are examined through detrending analysis. We found that both periods are multifractal and present self-similarity in the detrended probability density function (PDF). The Hurst exponent over short time intervals shifts from less than 0.5 ($\sim$ 0.42) in Period 1 to closer to 0.5 in Period 2 ($\sim$ 0.49), indicating that the market has gained efficiency over time. 
\end{abstract}

%\begin{keyword}
%% keywords here, in the form: keyword \sep keyword

%% PACS codes here, in the form: \PACS code \sep code

%% MSC codes here, in the form: \MSC code \sep code
%% or \MSC[2008] code \sep code (2000 is the default)
%Bitcoin, Stylized facts, Probability distribution function, Detrended Fluctuation Analysis, Hurst exponent
%\end{keyword}

%\end{frontmatter}

%%
%% Start line numbering here if you want
%%
% \linenumbers

%% main text
\section{Introduction}\label{sec:intro}

% definition of money and crypto, and the difference
Cryptocurrencies offer a decentralized and innovative alternative to traditional financial systems~\cite{Corbet2019Cryptocurrencies}. Although not globally accepted as a digital currency, many consider cryptocurrencies valuable assets to store and retrieve value when needed \cite{Crosby2016Blockchain}. With more than 2,000 distinct cryptocurrencies in circulation, their potential and prominence continue growing~\cite{Noam2021Macro-Economics}. In this dynamically evolving market, Bitcoin stands out as the dominant one and historically the best-performing cryptocurrency due to its price volatility and sustained growth \cite{panyagometh2024effect}. Bitcoin was introduced by Nakamoto \cite{Nakamoto2008} in 2008 as the first decentralized cryptocurrency, Bitcoin has seen a substantial increase in its market value. CoinMarketCap statistics reveal that Bitcoin's market capitalization reached over trillion USD in the first quarter of 2025, representing 48.09\% of the aggregate market capitalization of the main crypto-assets \cite{CoinMarketCap}. Previous analyses display a low correlation between Bitcoin
and global equity factors before 2020 but increased significantly afterward, particularly with technology-related assets, while remaining relatively weak with financial indicators \cite{che2023crypto}, except for S\&P500 \cite{nguyen2022correlation,grobys2023fractal}, especially during periods of high volatility or uncertainty. One possible explanation is Bitcoin's operability as an open, accessible network, not bound by conventional trading hours, along with a sustained increase in institutional investor participation, making it more sensitive to global financial factors. 

With an increasing number of users adopting Bitcoin despite its significant fluctuations, high volatility, and dependence on evolving technology, this emerging asset continues to gain importance in our economy. Previous studies have extensively explored various aspects of Bitcoin, including its market efficiency \cite{Urquhart2016Inefficiency, Kurihara2017market, Jiang2018Time_varying, Caporale2018Persistence, Noda2021evolution, Bariviera2017, Mnif2020cryptocurrency, Mnif2021COVID, Fernandes2022resilience, Montasser2022COVID, Kakinaka2022Cryptocurrency}, regulatory requirements \cite{Brito2014Bitcoin, Bohme2015Bitcoin, Hendrickson2017Banning, Marian2013Cryptocurrencies, Gross2017taxation}, market dynamics, and its correlations with various asset classes \cite{Conrad2018Long, Ciaian2018Virtual, Klein2018Bitcoin, Eross2019Intraday, Ariya2020CryptocurrenciesMarkets, arias2021methods, Hu2019Cryptocurrencies, Filho2021Leverage, Cremaschini2023stylized, Ghosh2023Return, Bariviera2017, Filho2021Leverage, Wen2022Intraday, Zhang2018stylized}, with the intention of quantifying Bitcoin price fluctuations and evaluating their proximity to the mechanism of market prices. Given that the Bitcoin pricing mechanism differs from conventional stock market price indexes, we propose the analysis by examining robust patterns or \textit{`stylized facts'} of Bitcoin price fluctuations, which can enhance our understanding of market efficiency. This analysis is based on `tick-by-tick' recorded data from Reuters Datascope. The motivation lies in unveiling precise time-scale properties of Bitcoin time series, allowing comparisons with the widely acknowledged stylized facts from stock market data , e.g. fat tails, volatility clustering, short-time correlations, and self-similarity measurements. To do this, we analyze the intraday data for the BTC / USD exchange, spanning the years 2018 to 2022. The data set has a frequency of 10 minutes and trades occur continuously throughout the 24-hour day.\par %We go beyond well-known properties found in financial markets and their instruments; our goal is to incorporate these outputs into a previously defined model for stock market price indices and evaluate its effectiveness.
The paper is organized as follows: Section \ref{sec:SF} recalls well-known stylized facts observed in financial time series. Section \ref{sec:background} presents a background of Bitcoin, which is essential to understand the dynamics of Bitcoin transactions. In Section \ref{sec:governing}, we provide the governing equations adapted from our closest model obtained for stock market prices for this study, along with the notation used in this paper. In Section \ref{sec:facts}, we divide the Bitcoin time series into different time periods characterized by the abrupt change in volatility, and then we examine the stylized facts for each period, respectively. Finally, we provide a summary of our findings based on the comparative analysis of the S\&P500 and Bitcoin.

\section{Stylized facts of financial markets} \label{sec:SF}
For the evaluation of financial models and the development of econometric theories, researchers focus on studying persistent statistical characteristics of the market, commonly referred to as \textit{`stylized facts'}.

Stylized fact is a widely adopted concept in economics, referring to the common statistical properties observed across datasets spanning different timeframes. The concept of stylized facts, originally introduced in macroeconomics to describe statistical patterns characterizing macroeconomic growth over extended periods and across diverse countries \cite{Kaldor1961Capital}, has undergone extensive examination in traditional financial instruments. In stock markets, widely recognized stylized facts include phenomena such as fat-tailed distribution, volatility clustering, self-similarity of market returns, and seasonality in time series \cite{Bollerslev1994ARCH, Shakeel2018Stylized,arias2021methods}. The analysis of stylized facts is crucial to establish the foundation for the development of theoretical forecasting models (\cite{Ryden1998Stylized, Liesenfeld2000Stochastic, Andersson2001Normal, Carnero2004Persistence, Bulla2006Stylized, Malmsten2010Stylized}). 
A substantial body of evidence suggests that cryptocurrency markets exhibit the key stylized characteristics observed in foreign exchange markets \cite{segnon2020forecasting, Bariviera2017}. These findings motivate a careful analysis of the most well-known stylized facts in stock markets before seeking future modeling approaches. In this section, we dedicate ourselves to recalling each of these well-known stylized facts along with their corresponding descriptions, as listed below. 

\begin{enumerate}
    \item \textbf{Fat-tailed distribution of returns}: This phenomenon was initially recognized by Mandelbrot \cite{Mandelbrot1963}, based on empirical distributions of financial returns (and logarithmic returns) that exhibit heavy-tailed distributions, deviating from the Gaussian distribution. The presence of heavy tails indicates a higher likelihood of extreme events than predicted by a normal distribution \cite{Stoyanov2011FatTailed}. Within a self-similar fat-tailed distribution, the tails can be characterized by a power law relation denoted $P(x,t) \sim t^{-H}F(xt^{-H})$, where $P(x,t)$ represents the probability distribution function (PDF) of the price return $x$. $F$ was widely assumed to follow a distribution known as a Levy distribution from the 1990s, $F(x)=L_{\alpha}(x)$, $H=1/\alpha$  where $\alpha \in (0,2]$ \cite{Levy1937, Mantegna1994Stochastic,Mantegna1999Econophysics}. After 2000, $F$ was considered as a q-Gaussian distribution, $F(x)=g_{q}(x)$, $H=1/\alpha$ , where $\alpha=\frac{3-q}{\xi}$ with $q \in (1,3)$  and $\xi$ as constant parameter\cite{Borland1998Microscopic,Malamud2003Tails, alonso2019q}. More recently $F$ has been viewed as the solution of a general porous media equation solved through local derivatives $H(t)=1/\alpha(t)$, where $\alpha(t)=\frac{3-q(t)}{\xi(t)}$ \cite{gharari2021space,tang2023variable}.
    %Originally, the description of the fat-tailed distribution of price returns was attributed to a L\'evy-stable distribution \cite{Mandelbrot1963}, which was later refined as a truncated L\'evy flight,. Subsequent investigations found that the $q$-Gaussian distribution offers an improved description of return distributions \cite{Tsallis1988Possible, Borland1998Microscopic, Tsallis2003Nonextensive}, with the tail conforming to \Blue{$P(x,t) \sim x^{\alpha}$}
    
    \item \textbf{Short-time autocorrelation of returns}: The autocorrelation function (ACF) quantifies the relationship between the current data and historical data, showing the extent to which there is some form of `memory' on the market \cite{Vilela2013Fluctuation, Nounou2000Multiscale}. Memory is commonly defined as the critic's shifted time for the autocorrelation process $\tau_c=1/ACF(0) \int_{0} ^{\infty} ACF(\tau) d\tau$ \cite{kasdin1995discrete}. In terms of financial data analysis, ACF is evaluated using price returns and describes a rapid decay during very small intraday time scales, before decaying to zero value. This trend aligns with the efficient market hypothesis \cite{Filho2021Leverage}, which signifies finite memory; it retains correlation with values closer back in time only.
    
    \item \textbf{Volatility clustering}: Volatility in finance measures the extent to which actual returns deviate from average returns over a designated time span \cite{Kumar2016Risk}. It is related to asset risk management, and higher volatility implies a greater chance of substantial losses~\cite{DeSilva2017Diversification}. Volatility is commonly quantified as the standard deviation ($\sigma$) of returns, and it plays an important role as an estimation tool in empirical investigations in financial studies, especially when examining new and emerging assets \cite{Ghosh2023Return}. With respect to stylized facts, volatility clustering is a widely observed phenomenon, particularly in speculative return time series. Mandelbrot initially identified this pattern \cite{Mandelbrot1963}, denoting the tendency for substantial changes to succeed with similarly substantial changes, and minor changes to be succeeded by comparably minor changes. This underscores the empirical reality that price fluctuations are nonstationary processes as the fluctuations are not identically distributed and the distributions experience temporal shifts \cite{arias2021methods}. 

    \item \textbf{Self-similarity and fractality}: Broadly defined, a fractal refers to a geometric shape that exhibits fragmented characteristics, with each fragment (at least approximately) resembling a reduced-scale replica of the whole structure \cite{Chen2016Finite-size}. A fractal can be described by its \textit{fractal dimension} $D_f$ \cite{Mandelbrot1982fractal}, which is obtained-- in the box-counting scheme-- using the scale-invariant relation between the number boxes needed to cover the whole object and the linear size of the boxes. In the self-similar time series, it quantifies how \textit{rough}, \textit{irregular} or \textit{complex} the time series is over different time scales, and is directly related to the Hurst exponent, $H$ via the relation $D_f=2-H$. Fractals have been studied extensively in finance and economics, including the stock market \cite{Lee2017Asymmetric}, gold markets~\cite{Mali2014Multifractal}, electricity price periods~\cite{Wang2013Multifractal}, crude oil markets~\cite{Gu2010Multifractal}, and the shipping markets \cite{Chen2016Finite-size}. In the domains of finance and economics, substantial research has explored the concept of fractals and scaling laws \cite{Onalan2004Self-similarity}. Scaling laws establish a connection between price returns computed over various sampling intervals, highlighting that the shape of the PDF of the price return remains consistent as the time scale varies $P(x,t) \sim t^{-H}F(xt^{-H})$\cite{gharari2021space}, where a self-similar time series denotes $\sqrt{\langle x^2 \rangle} \propto t^{H}$. Fractal analysis techniques such as rescaled range analysis (R/S) and Detrended Fluctuation Analysis (DFA) have been pivotal in these investigations \cite{Lahmiri2018Chaos}. Both methods yield a spectrum of the Hurst exponent, enabling the determination of the fractality of the time series. However, the R/S statistic is susceptible to outlier influence, which could lead to a biased estimate of the Hurst exponent \cite{Gu2010Multifractal}. Therefore, in this study, we adopt the multifractal detrended fluctuation analysis (MF-DFA) method, which is a generalization of the DFA, due to its proficiency in handling non-stationary time series \cite{Lahmiri2018Chaos}. 
\end{enumerate}

In addition to these well-acknowledged facts, financial markets exhibit other widely observed statistical characteristics, one of which is the anomalous diffusion at the peak of the PDF of returns. Bachelier's original work \cite{Bachelier1900} on market fluctuations also introduced the first option pricing model grounded in a Brownian motion, associating this motion with price-return fluctuations. Later, the Black-Scholes equation was proposed to improve the Brownian motion model to prevent the possibility of negative prices \cite{Black1973Pricing}. The Black-Scholes model is alternately referred to as the geometric Brownian model. However, these models fall short of effectively characterizing numerous dynamic processes \cite{Weron2009Anomalous}. This limitation arises from the fact that the mean square displacement of the price return scales with time exhibiting a fractional exponent deviating from conventional Brownian motion \cite{Sposini2022Towards}, $H \neq 0.5$ and is not a unique value $H(t)=1/\alpha(t)$; this deviation is known as anomalous diffusion. Anomalous diffusion is widely observed within financial data and is expressible through a power law relation $P_{max} \sim t^{-1/\alpha(t)}$, where $P_{max}$ represents the peak of the PDF of the future price return, and $t$ represents the predicted time \cite{alonso2019q}. Here, $\alpha=2$ indicates normal (Brownian) diffusion, $\alpha<2$ implies superdiffusion, and $\alpha>2$ denotes subdiffusion \cite{Sposini2022Towards}. Moreover, the value of $\alpha$ is intricately tied to the Hurst exponent ($H$) through the relation $H=1/\alpha(t)$ \cite{gharari2021space, Sposini2022Towards, Ghosh2023Return}. For example, this relationship is valid in fraction Brownian motion, which is a self-similar Gaussian stochastic process characterized by stationary power-law correlated increments \cite{gharari2021space}.

The aforementioned statistical properties are typically observed in traditional markets. This paper seeks to evaluate these properties in the context of Bitcoin, drawing comparisons with conventional markets. Given the distinctive framework of Bitcoin compared to traditional markets, the following section will offer a brief overview of its characteristics.

\section{Bitcoin's Operational Framework}\label{sec:background}
%1. Network; 2. transaction; 3. blockchain; 4. mining; 5. institutions; \\
Cryptocurrencies are decentralized digital assets that rely on encryption to verify transactions. 
Bitcoin, being the first, largest, and most well-known cryptocurrency, operates differently from centralized money systems. For centralized money, currency is issued by central banks, and transactions involve intermediaries like banks. In contrast, Bitcoin employs a decentralized digital infrastructure called ``blockchain" for peer-to-peer transactions and value storage. This allows Bitcoin to operate independently of any government, company, or financial institution. The concept of blockchain was initially proposed in 1991 \cite{Haber1991Time-Stamp} and was further developed by Satoshi Nakamoto \cite{Nakamoto2008}. A blockchain is a public ledger of all transactions and digital events shared among participating parties since the creation of Bitcoin \cite{Underwood2016Blockchain, Crosby2016Blockchain}. The blockchain maintains a certain and verifiable record of every transaction and ownership and once logged, the transaction can never be erased \cite{Crosby2016Blockchain, Vujicic2018Blockchain}. Furthermore, blockchain technology prevents double spending, improving the security level of the Bitcoin cryptocurrency \cite{Chuen2018Cryptocurrency}. This technology creates a network of traders, and each node in the network retains a copy of the ledger record \cite{Vujicic2018Blockchain}. 

Transferring Bitcoin from one owner to another is based on blockchain technology, where Bitcoin owners execute transactions using their private and public keys. The private key serves as a digital signature, while the public key is used to verify transactions. During a transaction, Bitcoins are sent from one address to another, requiring the sender's private key for validation. Transaction details must include the Bitcoin address, the amount to be transferred, and the public key of the next owner \cite{Vujicic2018Blockchain}. Subsequently, the transaction is broadcast to the Bitcoin network, rapidly disseminating across connected nodes to notify them of its occurrence. However, before being added to the blockchain, this transaction undergoes a validation process known as Bitcoin mining. Miners engage in a competition, a `proof-of-work puzzle, to include the transaction in a block. The winning miner then transmits the block to all nodes, allowing them to expand and update their copies of the blockchain \cite{Zaghloul2020Bitcoin}. Only miners possess the ability to add a transaction to the digital ledger, transitioning it from pending to confirmed. 
%, and the efficiency was gradually improved until Nakamoto implemented the first real blockchain and used it as the core technology for the Bitcoin cryptocurrency system \cite{Zaghloul2020Bitcoin}.    \\
%In his white paper, Satoshi Nakamoto \cite{Nakamoto2008} described the peer-to-peer version of electronic cash that would allow online payments to be sent directly from one party to another without going through a financial institution. Bitcoin was introduced as the first realization of this concept in an open-source program implementing this new protocol \cite{Crosby2016Blockchain}.
%Bitcoin is proposed to feature a distributed timestamp server, designed to function as a computational proof generator for verifying the order sequence of transactions \cite{Vujicic2018Blockchain}. Bitcoin owners have a private key and a public key. The private key is used as a digital signature, and the public key is used for verifying transactions. Each individual transaction is characterized by using not only a collection of digitally signed hashes from the previous transaction data, but also the public key of the next owner \cite{Vujicic2018Blockchain}.  \\
%In the Bitcoin network, a node is a network entity that is connected to one or multiple other similar nodes, and the directly connected nodes are referred to as peers \cite{Zaghloul2020Bitcoin}. Therefore, the peer-to-peer transaction refers to the transaction occurring between two individuals directly. \

Miners play a crucial role in Bitcoin systems, assembling a special transaction known as a coinbase transaction, which relates to the creation of new coins in the network.
Bitcoin operates independently of a centralized authority responsible for issuing new coins, and instead, the coins are automatically generated by the Bitcoin blockchain system as a reward for the first miner who successfully completes the proof-of-work consensus mechanism \cite{Nakamoto2008}. The initiation of a new coin is triggered by the first transaction in a block, and the ownership of this coin is attributed to the creator of the block \cite{Nakamoto2008}. This incentive structure motivates miners to validate transactions, thereby injecting coins into circulation. The mining process involves solving a mathematical puzzle, and the miners are rewarded for finding the solution, resulting in the creation of a new block that is updated on the blockchain \cite{Tschorsch2016Bitcoin}. The design of the Bitcoin network aims to produce one block roughly every 10 minutes. As computational power increases, the network maintains a relatively stable block creation time by progressively increasing the difficulty of generating new blocks \cite{Vujicic2018Blockchain}. Although mining remains the exclusive means of generating new coins, it comes with a significant energy consumption. Furthermore, individual miners may face a notably low likelihood of achieving successful returns. To address this, solo miners collaborate by uniting their mining capabilities and forming mining pools. Eventually, these pools evolve into substantial organizations with considerable computational prowess, allowing them to contend with other major entities. The rewards are then distributed among the participants based on their individual contributions \cite{Zaghloul2020Bitcoin}. 

One fundamental characteristic of Bitcoin is its deliberate limitation on coin supply. The initial Bitcoin coin reward was established at 50 coins, which is the total amount of Bitcoin when it was first created. This reward is designed to undergo halving every 210,000 blocks, which is approximately four years \cite{Vujicic2018Blockchain}. With a fixed total supply of 21 million Bitcoin, currently approximately 93\% of this total has been brought into existence. Even after the full issuance of coins, Bitcoin will remain exchangeable among owners, with rewards shifting to the verification of transactions.

Bitcoin offers a 24/7 market accessible worldwide. Its unique attributes enable transactions to occur at a significantly accelerated pace compared to fiat currency, leading to a notably more volatile market compared to conventional financial markets.

%==================================================================================================

\section{Governing equations}\label{sec:governing}
%In this section, we present the theoretical background of this study. We start by defining the variables and presenting the governing stochastic differential equations. Then we derive the expression of the probability distribution function in terms of Katugampola derivative. The calculation of the anomalous diffusion of the time series is illustrated with respect to the second moment. Finally, the expression for the autocorrelation is presented in an integral form. 
A common approach for analyzing option pricing is related to the well-known Black-Scholes equation (BSE) as the governing equation for modeling price return. We recall our governing equation from \cite{Arias-Calluari2022stationarity} which is used to model stock market indexes. This equation allows prices to fluctuate with a trend and a stochastic noise. This stochastic noise can be well described by a $q$-Gaussian distribution. The time evolution of the PDF of stochastic processes can be presented using a Fokker-Planck equation (FPE). In our analysis, we use fractional FPE to describe the time evolution of the PDF of the price return, and use the local Katugampola fractional operators for the fractional $q$-Gaussian process. The anomalous diffusion is defined in terms of the second moment of the PDF. The autocorrelation is presented in an integral form. 

\begin{enumerate}
\item \textbf{Governing equations of the stock market}: The market index of any given stock market data is denoted as $I(t)$, and the simple price return in a time interval from current time $t_0$ to future time $t$ is defined as
\begin{equation} \label{eq:simple_price_return}
    X(t_0, t) = I(t_0+t) - I(t_0). 
\end{equation}
The stock market index fluctuates over time in a random process. In this analysis, it is assumed that the stock market index $I(t)$ can be decomposed to a deterministic trend $\Bar{I}(t)$ and a stochastic noise $\hat{I}(t)$
\begin{equation} 
    I(t) = \overline{I}(t) + \hat{I}(t). 
\label{eq:detrend_I}
\end{equation}
The price return $X(t)$ can also be divided into two parts: a deterministic component $\Bar{X}(t)$ and a stochastic $q$-Gaussian noise $x(t)$
\begin{equation} 
    X(t) = \overline{X}(t) + x(t),
\label{eq:trended_x}
\end{equation}
where $\overline{X}(t)= \overline{I}(t_0+t)-\overline{I}(t_0)$, and $x(t)= \hat{I}(t_0+t)-\hat{I}(t_0)$.
The increment of the price return is calculated by the difference between the consecutive points of the price return, written as
\begin{equation}
\begin{split}
    &X^*(t)=I(t+1)-I(t) \\
    &x^*(t)=\hat{I}(t+1)-\hat{I}(t)
    \end{split} .
    \label{eq:increment_price_return}
\end{equation}
From here onward, $t$ represents the normalized time obtained by $t= time/\Delta t$, where $time$ is the time in minutes and $\Delta t = 10 \, mins$ is the frequency of the Bitcoin price index.

The main interest in financial market forecasting is in predicting the future price return $t_0+t$. In this context, we employ a probability approach, where we assume that the simple price return $X$ is a random variable characterized by a time-dependent PDF $P(x,t)$. Then our primary goal is to establish the governing equation that describes the evolution of $P(x,t)$.

Eq.~\ref{eq:trended_x} can be expressed as a stochastic derivative that is different from the standard derivative, where the quantity $x(t)$ is stochastic and, therefore, not differentiable, written as
\begin{equation} 
    \frac{dX}{dt} = \frac{d\Bar{X}}{dt} + \frac{dx}{dt}. 
\label{eq:detrend_x}
\end{equation}
That is conveniently written as from a stochastic It\^o-Langevin equation of the form
\begin{equation}
    \frac{dX}{dt}=\mu(X,t)+\sigma(X,t)\eta(t),
    \label{eq:SDE}
\end{equation}
where $\mu$ is the trend of the time series, $\sigma$ is the volatility, $\eta(t)$ is a white noise with $<\eta>=0$.
Using the It\^o's lemma \cite{Ito1951}, it is possible to derive the partial differential equation that describes the temporal evolution of the probability density function $P(X,t)$ of the stochastic variable $X(t)$, presented as
\begin{equation}
    \frac{\partial P}{\partial t}= -\frac{\partial}{\partial X} (\mu(X,t) P)+\frac{1}{2} \frac{\partial^2}{\partial X^2}(\sigma^2(X,t)P).
\end{equation}
For a detrended price return, $\mu(X,t)=0$, and the evolution of PDF is reduced to
\begin{equation}
    \frac{\partial P(x,t)}{\partial t}=
    \frac{1}{2} \frac{\partial^2}{\partial x^2}( \sigma^2(x,t)P(x,t)).
    \label{eq:difussion}
\end{equation}
The work of Borland suggests that the volatility depends on the PDF so that the diffusion process becomes non-linear \cite{Borland1998Microscopic}. Based on our earlier empirical analysis of the S\&P500 stock market index, we have found that \cite{alonso2019q}
\begin{equation}
    \frac{\partial P(x,t)}{\partial \tau}
    =D_0\frac{\partial^2 P^{2-q}(x,t)}{\partial x^2},
    \label{eq:master}
\end{equation}
where $D_0$ is a diffusion constant, $\tau = t^\xi$, being $0<\xi\le 1$ an empirical exponent obtained from the empirical data. The solution of Eq. \ref{eq:master} with the Dirac's delta as initial condition $P(x,t=0)=\delta(x)$ is given by \cite{alonso2019q}
\begin{equation}
 P(x,t)=\dfrac{1}{(D_0 t)^{H}} \left[ g_q \left(\dfrac {x}{(D_0 t)^{H}}\right) \right],
\label{eq:pdf_q-Gaus}
\end{equation}
where $H=1/\alpha=\xi/(3-q)$ and $g_q(x)$ is the normalized $q$-Gaussian distribution distribution defined by
\begin{equation}
    g_q(x)=\dfrac{1}{C_{q}} \left[1-(1-q)x^2\right]^{\dfrac{1}{1-q}}.
    \label{eq:g_q}
\end{equation}
This is a generalization of the Gaussian distribution; if $q=1$ the Gaussian distribution is recovered. The normalization constant $C_{q}$ for $1<q<3$ is given by
\begin{equation}
C_q =\sqrt{\dfrac{\pi}{q-1}} \dfrac{\Gamma((3-q)/(2(q-1)) ) }{\Gamma(1/(q-1)) }.
\label{eq:Cq}
\end{equation} 

\item \textbf{Fractional diffusion equations}: Now we present the diffusion equation in fractional form. Eq.~\ref{eq:master} can be rewritten in two different formats. The first case is to convert it into a Fokker-Planck equation with time-dependent volatility. This is performed by using the rule $d\tau =t^{\xi-1} dt $ to convert it into
\begin{equation}
    \frac{\partial P(x,t)}{\partial t}
    =D_0 t^{\xi-1}\frac{\partial P^{2-q}(x,t)}{\partial x^2},
    \label{eq:master_FPE}
\end{equation}
and by comparing Eq. \ref{eq:difussion} and \ref{eq:master}, we obtain the expression for the volatility as
\begin{equation}
    \sigma(x,t)=\sqrt{2D_0}t^\frac{1-\xi}{2}P^{\frac{1-q}{2}}(x,t).
\end{equation}
In a more recent work \cite{gharari2021space}, it has been proposed that Eq.~\ref{eq:master} can be written as a local fractional differential equation with the Katugampola fractional operator. The Katugampola fractional derivative for $0<\xi\le 1$ is defined by
\begin{equation}
\frac{d^\xi f}{d^\xi t}:=\lim_{\epsilon \to 0}\dfrac{f(te^{\epsilon t^{-\xi}})-f(t)}{\epsilon}.
\label{Eq:Katugampola}
\end{equation}
The Katugampola fractional derivative has the property of
\begin{equation}
\frac{d^\xi f}{d^\xi t}=t^{1-\xi}\frac{df}{dt}.
\label{Eq:Katu_property}
\end{equation}
Using the above properties, we can derive the fractional form of Eq.~\ref{eq:master} as
\begin{equation}
    \frac{\partial^\xi P(x,t)}{\partial^\xi t}
    =D_0\frac{\partial^2 P^{2-q}(x,t)}{\partial x^2}.
    \label{eq:fractional}
\end{equation}
This proves that Eqs.~\ref{eq:master_FPE} and \ref{eq:fractional} are the same equations. 

\item \textbf{Anomalous diffusion}: In the following, we connect the expression for anomalous diffusion with the second moment of the time series. For a wide range of diffusion processes, the PDF follows the scaling solution and can be written in a generalized form as \cite{tang2023variable}
\begin{equation}
    P(x,t)= \frac{1}{\phi(t)} F \left[ \frac{x}{\phi(t)} \right],
    \label{eq:scaling}
\end{equation}
where $F$ is a universal function generally, being the normalized $q$-Gaussian distribution function in this study. 
The anomalous diffusion is normally recognized and characterized using the second moment of the random variable $x$ as a function of time, i.e.
\begin{equation}
\begin{split}
\left\langle x^2\right\rangle  &\equiv \int_{-\infty}^{\infty} x^2 P(x,t) \,dx \\
&= \int_{-\infty}^{\infty}\frac{x^2}{\phi(t)}  F \left[ \frac{x}{\phi(t)} \right] \,dx,
\end{split}
\end{equation}
which casts to ($\frac{x}{\phi(t)}=y$)
\begin{equation}
    \begin{split}
        \left\langle x^2\right\rangle &= \phi(t)^2 \int_{-\infty}^{\infty} y^2 F(y) \, dy \\
        &= \phi(t)^2 \sigma_0^2.
    \end{split}
\end{equation}
For a normalized distribution, $\phi(t)\propto t^{H}$, and $\sigma_0$ can be ignored, where $H$ is the \textit{Hurst exponent}, the leading quantity to characterize the time series. Therefore the anomalous diffusion is identified and characterized as
\begin{equation}
    \left\langle x^2\right\rangle = \phi(t)^2 \propto t^{2H}.
    \label{eq:moment}
\end{equation}  
The PDF is assumed to satisfy the scaling relation given by Eq.~\ref{eq:scaling}. Changing $x=0$ into Eq.~\ref{eq:scaling} we obtain $P(0,t)=F(0)/\phi(t)$. Assuming that the PDF peaks at $x=0$ (i.e. $P_{max}(t) = P(0,t)$ and using Eq.~\ref{eq:moment} the second moment is reduced to
\begin{equation}
    \sqrt{\left\langle x^2\right\rangle} \propto t^H\sim 1/P_{\text{max}}(t),
    %<x^2> = \phi(t)^2 \propto t^H\sim1/P_{max}(t).
    \label{eq:moment_reduce}
\end{equation} 
where $P_{\text{max}}(t)$ is the maximum of the PDF as a function of time. In other words, the exponent of the anomalous diffusion can be obtained from the time evolution of the peak of the PDF.

\item \textbf{Autocorrelation Function (ACF)}: The autocorrelation function can be derived from the SDE. For a fractional Brownian motion, the correlation function can be expressed as
\begin{equation}
\begin{split}
     E[X(t)X(t+\tau)] =  \int_{0}^{t+\tau}  \int_{0}^{t} W(t)W(t+\tau) dt d(t+\tau),
\end{split}
\end{equation}
where $W(t)=dX/dt$, and $\text{E}[...]$ is the expectation value. From Eq. \ref{eq:SDE}, we can represent the time series as
\begin{equation}
    X(t) = \int_{0}^{t} \left( \mu(X,t)+\sigma(X,t)\eta(t) \right) \, dt.
\end{equation}
Assuming the time series is detrended, we have $\mu(X,t)=0$. We can write the autocorrelation as
\begin{equation}
\begin{split}
    &\left< X(t)X(t+\tau) \right> \\ 
    &=  \left< \int_{0}^{t} \sigma(X,t') \eta(t') \, dt' \int_{0}^{t+\tau} \sigma(X,t'') \eta(t'') \, dt'' \right> \\
    &= \int_{0}^{t} \int_{0}^{t+\tau} \left<  \sigma(X,t') \eta(t')  \sigma(X,t'') \eta(t'')  \right> \, dt' dt''.
\end{split}
\end{equation}
From Eq. \ref{eq:SDE}, assuming $\sigma(x, t)$ and $\eta(t)$ are known, one can solve the integral to obtain the analytical expression of the autocorrelation.
\end{enumerate}
%==================================================================================================
\section{Stylized facts of Bitcoin market index}\label{sec:facts}
\begin{figure}[!htbp]
\centering
\includegraphics[scale=0.6, clip]{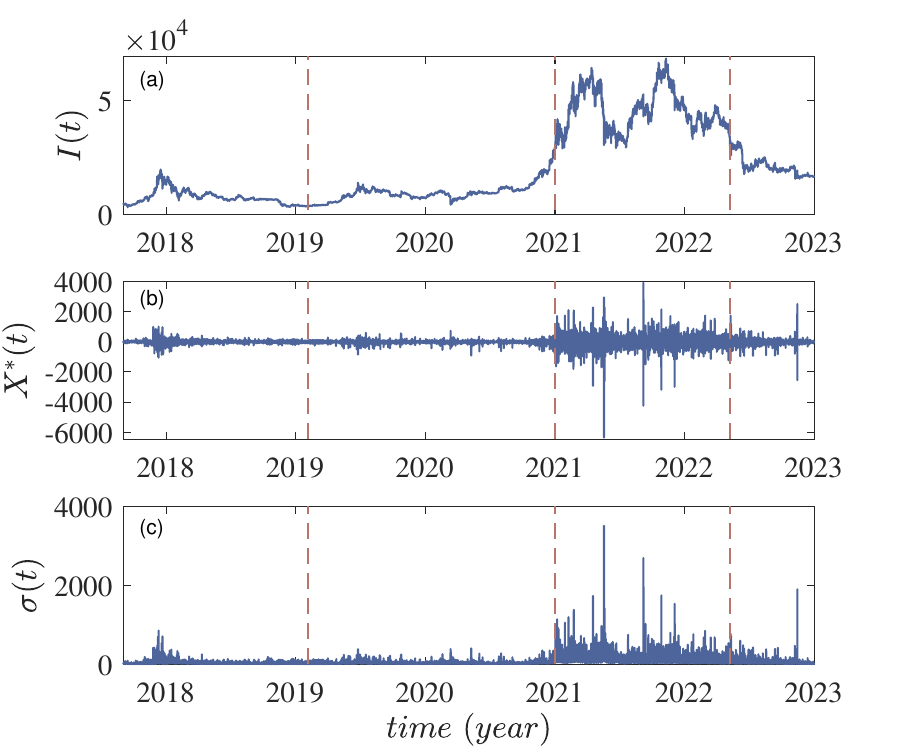}
\caption{ \textcolor{black}{(a) Bitcoin market index $I(t)$ from 03/09/2017 to 31/12/2022. (b) Simple price return $X^*(t)$ obtained from $\hat{I}(t)$ following Eq. \ref{eq:increment_price_return}. (c) The standard deviation $\sigma(t)$ of simple price return calculated by using a 1-hour moving window, where $\sigma(t)= \sqrt{\frac{1}{N-1} \sum_{t=1}^N (X^*(t)-\mu)^2}$, $X^*_i$ indicates a specific segment of the price return, and $\mu$ is the mean value of that segment.}}
\label{fig:Price_undetrend_std}     
\end{figure}

%\subsection{Distinctive periods in Bitcoin time series}
%In this study, we aim to explore the market characteristics of the Bitcoin market and to establish the stylized facts of Bitcoin. 
Financial time series exhibit diverse stylized facts over time. Consequently, segmenting the entire time series into distinct periods for detailed investigation becomes crucial. Our initial step involves identifying these different sections within the Bitcoin price index. We collected data on the Bitcoin price index from March 9, 2017, to December 31, 2022, with a 10-minute frequency. Using this dataset, we computed the simple price return $X^*(t)$ (Eq.~\ref{eq:increment_price_return}) and volatility $\sigma(t)$. The latter was calculated as the standard deviation of $X^*(t)$ on a rolling window of one hour. The results are shown in Figure \ref{fig:Price_undetrend_std} (a-c) for the price index $I(t)$, simple price return $X^*(t)$, and volatility $\sigma(t)$ respectively, providing a reference for segmenting the dataset. Data prior to April 2, 2019, were deemed spurious due to unrealistic jumps of the order of 100 USD and, therefore, excluded from further analysis. Figure \ref{fig:Price_undetrend_std}(a) illustrates the price index, revealing a substantial increase in price since 2021. This significant increase coincides with Elon Musk's purchase of \$1.5 billion worth of Bitcoin and Tesla's announcement to accept Bitcoin as payment. This marked a significant turning point, and aligns with previous multivariate time series analysis,which also recognize data from 2021 onward as a distinct period \cite{panyagometh2024effect,nguyen2022correlation}. Consequently, data points that span from April 2, 2019, to December 31, 2020, constitute an independent segment denoted as Period 1 in this study.
A notable shift in market dynamics unfolded between 2021 and May 2022. During this period, cryptocurrency prices surged significantly. However, a significant change occurred after May 2022. This period witnessed a series of crises that affected multiple cryptocurrencies and trading platforms. \Black{The collapse of LUNA from Terraform Labs resulted in a total loss of market value of over \$400 billion.} %Concurrently, the decision by Elon Musk to discontinue Bitcoin acceptance for Tesla car purchases contributed to an environment characterized by an increased frequency of substantial market downturns. 
Consequently, the data recorded after May 2022 warrant categorization as a distinct time period. Therefore, we labeled the data set from January 1, 2021, to May 9, 2022, as Period 2 in this study. Figure \ref{fig:Price_undetrend_std} (b\&c) depicts the simple price return and volatility, respectively, illustrating that Period 2 exhibits notably higher returns and volatility compared to Period 1.

%=======================================================================================
%=======================================================================================
%\section{Stylized facts of trended stock market time series}\label{sec:trended}
%In the preceding section, we delineated the discernible shifts in the characteristics of the Bitcoin price index, resulting in the division of our dataset into two distinct periods, namely Period 1 and Period 2. In this section, our focus turns to an exploration of the stylized facts within each of these periods, with a particular emphasis on the trended dataset.

%=======================================================================================
\subsection{Anomalous diffusion and fat-tailed distribution}\label{sec:anomalous}
Our analysis begins with an examination of the PDF of simple price return $X(t_0, t)$ as shown in Eq. \ref{eq:simple_price_return}. For both Period 1 and Period 2, we calculate the PDFs across various time intervals $t = t - t_0$, that is, we plot the relative diffusion time instead of the absolute time, ranging from the smallest diffusion interval $t = 1$ to the diffusion period of a year $t = 326 \, days$. The kernel density estimator, known for its ability to provide a smooth and accurate PDF estimation \cite{Weglarczyk2018}, is used with a kernel bandwidth set to 0.001. This ensures that the bandwidth is sufficiently small to capture the detailed structures of the PDFs. To compute the PDFs, we calculate the price return using Eq. \ref{eq:simple_price_return} for each time interval $t$ and then use the price return to calculate $P(X,t)$.

\begin{figure}[!ht]
\centering
\includegraphics[scale=0.6,clip]{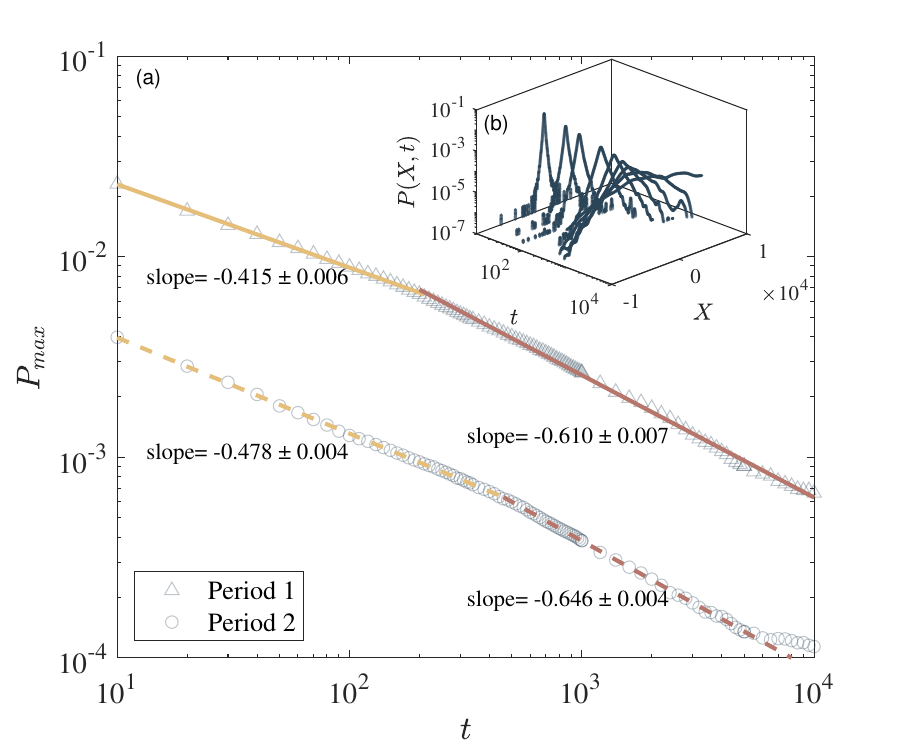}
\caption{(a) Time evolution of the peak of the PDF for Period 1 and 2 in the log-log scale shown in Black markers. Two well-defined slopes can be observed for each period. The colored lines show the fitted slope of the power-law relation. Both periods experience a transition from a weak subdiffusion regimen to a weak superdiffusion regime over time. (b) Time evolution of the PDF for Period 2 from 10 mins to 1000 mins. The PDFs present distinctive peaks and heavy tails at small time intervals and become flat as the time interval increases.}
\label{fig:peak_undetrend}     
\end{figure}

In Figure \ref{fig:peak_undetrend} (a), we present the peak of the PDF $P_{max}$ for Periods 1 and 2, respectively, in relation to time $t$. In particular, both time periods exhibit a power-law relationship between the peak of the PDF and time, expressed as $P_{max} \sim t^{-H}$. Linear curve fitting is applied to the data of Periods 1 and 2, respectively, to measure the power-law slope ($H=1/\alpha$). We note that a transition in the $H$ values is observed for both data sets, signifying a shift in the diffusion mode. For Period 1, the Hurst exponent $H$ is $0.415 \pm 0.006$ at small time intervals and $0.610 \pm 0.007$ at large time intervals, corresponding to the values of $\alpha=2.41 \pm 0.03$ and $\alpha=1.64 \pm 0.02$, respectively. For Period 2, the slope is $H=0.478 \pm 0.004$ at small time intervals and $H=0.646 \pm 0.004$ at large time intervals, corresponding to the respective values of $\alpha=2.09 \pm 0.01$ and $\alpha=1.54 \pm 0.01$. The anomalous diffusion exponent $\alpha$ is used to distinguish normal (Brownian, $\alpha=2, H=0.5$) from anomalous diffusion ($\alpha \neq 2, H \neq 0.5$). The super and subdiffusion regimes correspond to $\alpha>2, H<0.5$ and $0<\alpha<2, H>0.5$, respectively \cite{Sposini2022anomalous}. These results suggest that both periods of the Bitcoin time series undergo a transition from a weak subdiffusion regime to a weak superdiffusion regime over an extended period.

\begin{figure}[!ht]
\centering
\includegraphics[scale=0.6, clip]{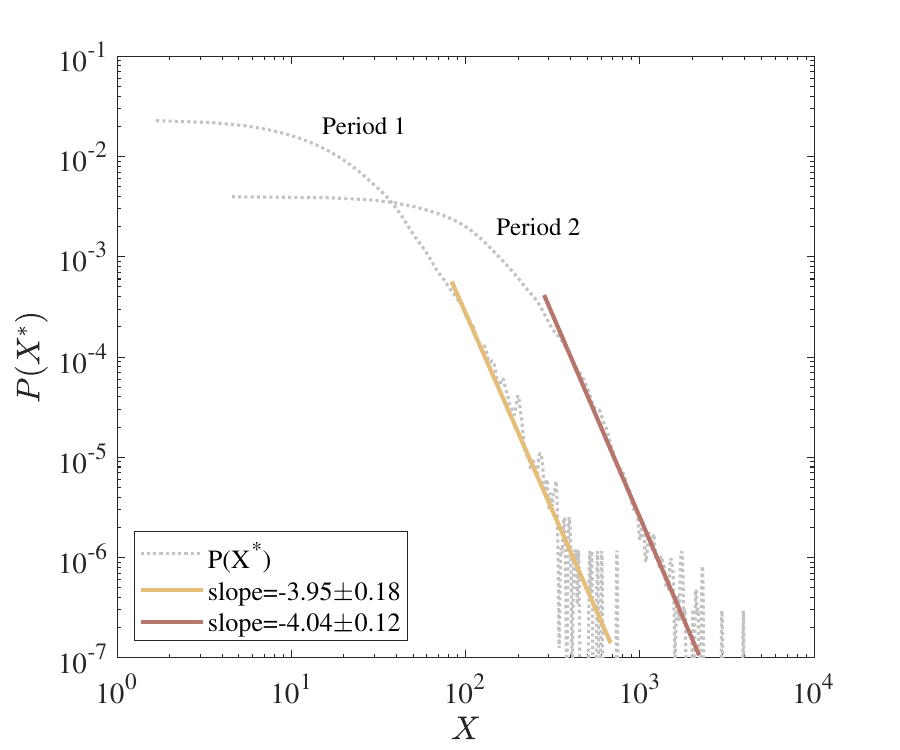}
\caption{ \textcolor{black}{Tail slope of PDF at the minimum time interval $t_0$ for Period 1 and 2 respectively in log-log scale. The dotted grey lines are the PDFs of price return, and the colored lines show the fitted slope of the tail for each time period. The fitted slopes show that the tail slope is outside the L\'evy regime, and fits to a $q$-Gaussian distribution.}}
\label{fig:tail_slope_section2_3}     
\end{figure}

Figure \ref{fig:peak_undetrend} (b) illustrates the time evolution of the PDF from 10 minutes to 1000 minutes (approximately 16.5 hours) of trading time. The PDF at the minimal time interval $t=1$ exhibits a heavy-tailed non-Gaussian distribution, gradually flattening and broadening as time progresses. 
To further explore the fat-tailed distribution of the PDF, it is essential to determine the tail slope of the PDF at the minimum time interval, $P(X^*, t=1)$. This slope, denoted as $\alpha$, is crucial for characterizing the type of distribution. The tail exponent for L\'evy distribution is calculated as $P(x,t) \sim x^{-(1+\alpha)}$, where $0<\alpha<2$ \cite{Levy1937, Mantegna1994Stochastic}. The exponent for the $q$-Gaussian distribution is $P(x,t) \sim x^\alpha$, where $\alpha=\frac{2}{1-q}$, and $1<q<3$ \cite{alonso2019q}.
In Figure \ref{fig:tail_slope_section2_3}, we present the calculated tail slopes of the PDF for each period. The slopes of the tails for Period 1 and Period 2 are $\alpha=3.95\pm0.18$ and $\alpha=4.04\pm0.12$ respectively. By comparing these values with the corresponding exponents of L\'evy and $q$-Gaussian distributions, we find that the slopes fall outside the L\'evy regime and instead fit well into the $q$-Gaussian regime. The values of $q$ calculated based on the tail slopes derived using the aforementioned relation are $q=1.51\pm0.02$ and $q=1.50\pm0.02$ for Periods 1 and 2, respectively. 

\begin{figure}[!ht]
\centering
\includegraphics[scale=0.6, clip]{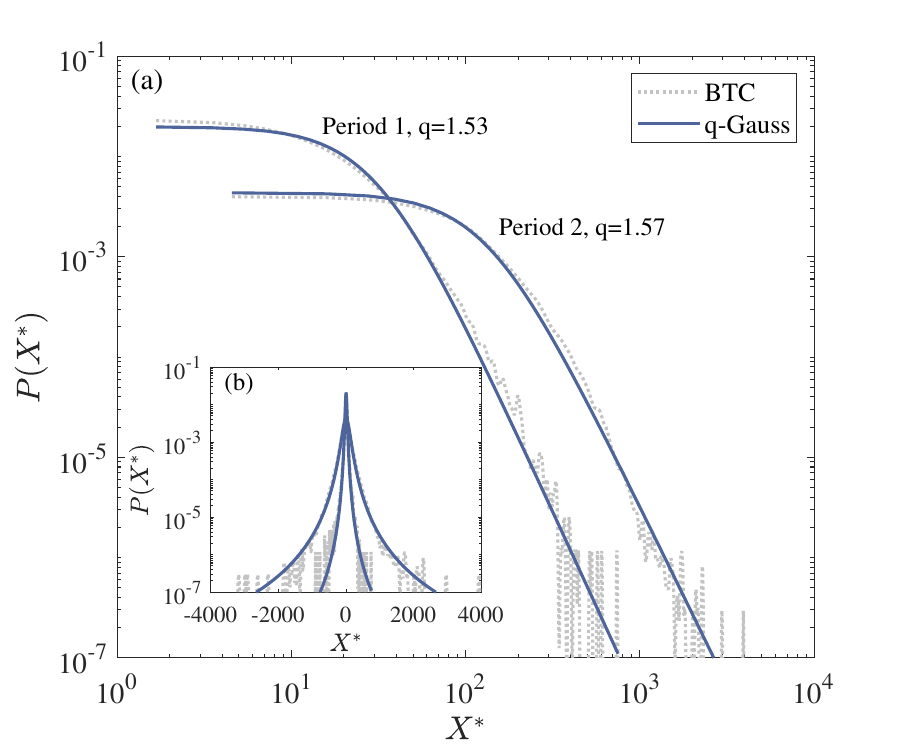}
\caption{$q$-Gaussian fitting conducted in semi-log scale for PDFs of simple price return at $t_0$ for both Period 1 and 2. The grey dotted curves are the PDFs and the Black curves are the fitted $q$-Gaussian distribution. (a) The right branch of the PDFs for Period 1 and 2 are plotted in log-log scale respectively. (b) The full PDF and the fitted distribution are plotted in a semi-log scale.}
\label{fig:qgaus_section2_3}     
\end{figure}

Upon examining the fat-tailed distribution, we found that the PDF in the minimum time interval ($t=1$) can be characterized by a $q$-Gaussian distribution. To substantiate this observation, we conducted a calibration to the $q$-Gaussian distribution for the PDFs corresponding to both Period 1 and Period 2. The results of this calibration process are shown in Figure \ref{fig:qgaus_section2_3}.
This calibration procedure was conducted on a semi-logarithmic scale by applying the relationship described in Eq. \ref{eq:pdf_q-Gaus}. This method involved taking the natural logarithm of the PDF and fitting it to the simple price return using a linear scale. Figure \ref{fig:qgaus_section2_3} (a) plots the right branch of the PDF using a logarithmic scale, and Figure \ref{fig:qgaus_section2_3} (b) illustrates both branches of the PDF in semi-logarithmic scale. In these figures, the gray dotted curves represent the PDF of the simple price return, while the black curves represent the fitted $q$-Gaussian distribution. In particular, the fitted distribution captures both the central and tail portions of the PDFs. The $q$ values derived from the semi-logarithmic fitting were determined as $q=1.53$ for Period 1 and $q=1.57$ for Period 2. Importantly, these values align with the $q$ values fitted from the power-law tail slope, affirming the consistency of the results that the diffusion is $q$-Gaussian. 
%The distribution exhibits a good fit in the central region of the PDF but shows discrepancies in the tails. Figure \ref{fig:qgaus_section2_3}(b) illustrates both branches of the PDF using a semi-log scale. The obtained values of $q$ for the $q$-Gaussian distribution were found to be $q=1.75$ and $q=1.59$ for Section 2 and Section 3 respectively. These $q$ results are inconsistent with the $q$ value calculated from the tail slope. The linear scale calibration primarily focuses on the central region of the PDF rather than the tail, leading to such discrepancies. To enhance the accuracy of the calibration, we took a further step by performing the $q$-Gaussian fitting in a semi-log scale. This involved applying a logarithm to the PDF and fitting it to the price return. The results of this semi-log fitting process are depicted in Figure \ref{fig:qgaus_section2_3}(c-d), which exhibit better fittings to the tail. The $q$ values obtained from the semi-log fitting were determined as 1.53 for Section 2 and 1.57 for Section 3, demonstrating a closer agreement with the $q$ values calculated from the tail slope. By comparing the fitted distributions in different scales, it becomes evident that the fitting in the semi-log scale yields more accurate results.

%=======================================================================================
\newpage
\subsection{Volatility clustering}\label{sec:volatility}
%\Red{Ghosh2023Return mention MF-DFA}
We used the autocorrelation function (ACF) to quantify volatility clustering in time series. The ACF is calculated on the incremental price return $X^*(t)$ using two methods: sample autocorrelation and chopping autocorrelation. The detailed definitions for each method are presented in the following.
%The autocorrelation function quantifies the extent of the correlation between the same variable measured at different time intervals. It measures the relationship between the value of a variable at a specific time lag and its original value within the time series. 

The time lag of the autocorrelation is denoted as $s$, representing real-time intervals in minutes for this study. For each $s$, the sample autocorrelation is defined as
\begin{equation}
    C(s)=\frac{\sum_{t=1}^{n-s} (X^*_{t+s}-\bar{X^*})(X^*_{t}-\bar{X^*})}{\sum_{t=1}^{n} (X^*_{t}-\bar{X^*})^2},
\end{equation}
where $X^*$ is the simple price return, $X^*_{t+s}$ is the price return shifted by $s$ minutes, and $\bar{X^*}$ is the mean value of the price return, calculated as
\begin{equation*}
    \bar{X^*}=\frac{1}{n} \sum_{t=1}^{n} X^*_t.
\end{equation*} 

\begin{figure*}[!hbt]
\centering
\includegraphics[scale=0.55, clip]{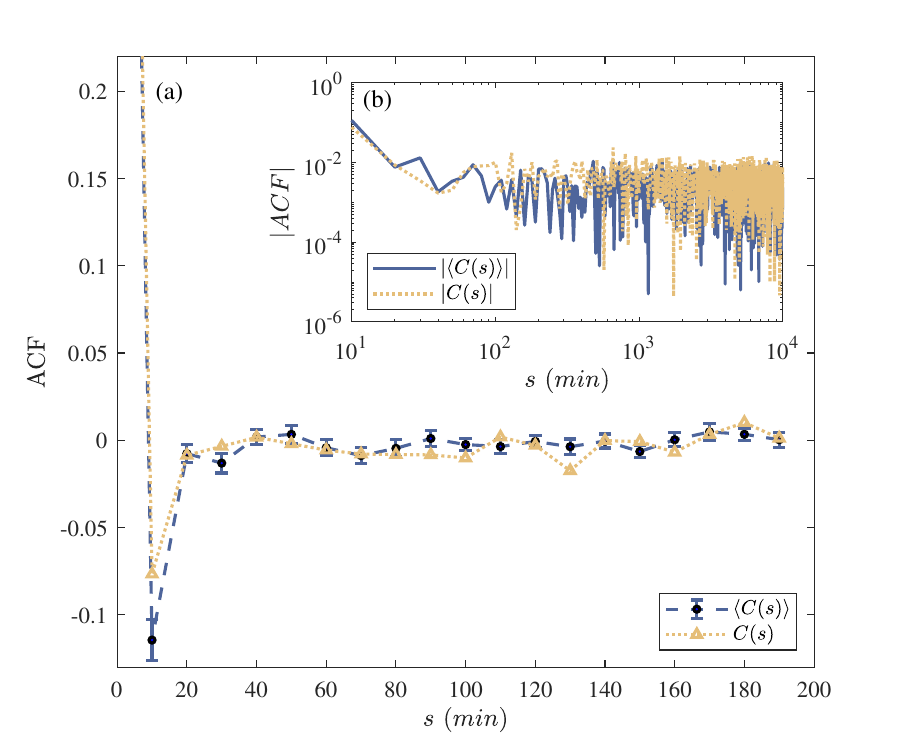}
\includegraphics[scale=0.55, clip]{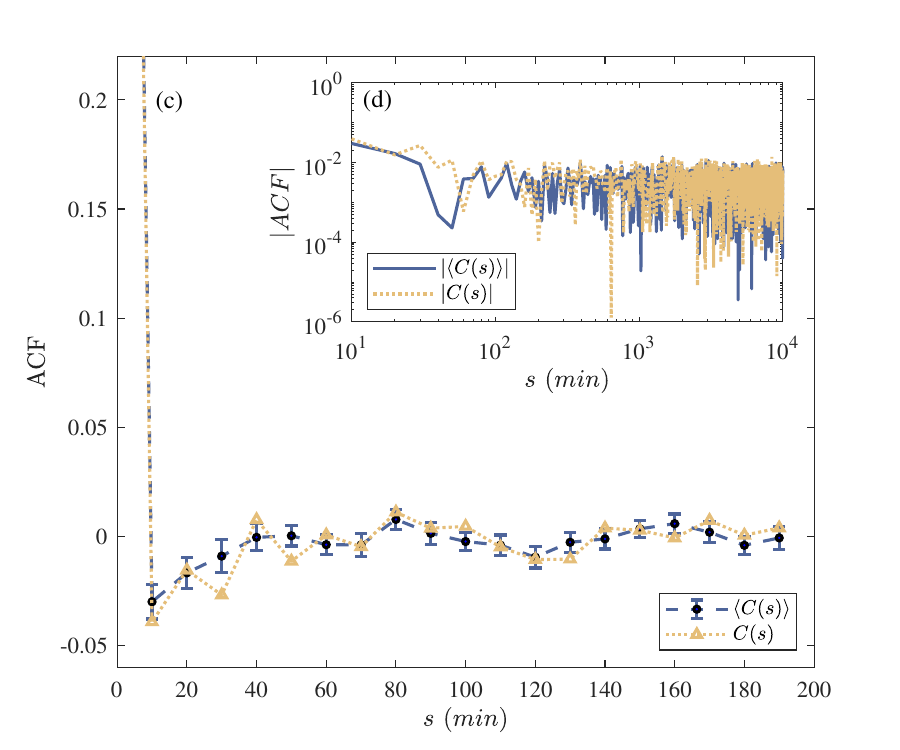}
\caption{Sample and chopping autocorrelation with fitting for Periods 1 and 2. For both periods, anti-correlation is observed for short times, and weak long-term autocorrelation is presented. (a) For Period 1, chopping ACF is plotted in blue, and sample ACF is plotted in yellow. Anti-correlation is observed in short times with periodicity.
% The solid red line shows the fitting of the ACFs. Fitting parameters are $a=-1.17, b=0.21, c=1.76$. 
(b) Absolute sample and chopping ACF of Period 1 plotted in log-log scale, representing a power-law relation. A linear fitting in the log-log scale measures the slope of the power-law relation as -1.17. (c) For Period 2, chopping ACF is plotted in blue, and sample ACF is plotted in yellow. Anti-correlation is also observed in short times with periodicity. 
%The solid red line shows the fitting of the ACFs. Fitting parameters are $a=-1.07, b=0.13, c=0.40$. 
(d) Absolute sample and chopping ACF of Period 2 plotted in log-log scale, with a power-law relation at initial times. A linear fitting in the log-log scale measures the slope of the power-law relation as -1.07. }
\label{fig:autocor}
\end{figure*}

We also computed the chopping autocorrelation of the price return for both periods, following the concept of calculating the ensemble autocorrelation. In the context of stochastic processes, the ensemble represents the statistical population of the process, where each member of the ensemble is a possible realization of the process \cite{McCauley2008ensemble, Lahiri2016optical}. For finance data where only a singular historical time series exists, constructing this ensemble involves decomposing the time series into an ensemble of subintervals of the data. Here, each trading day effectively constitutes one realization of the dataset, given the recurring nature of market statistics on a daily basis \cite{Bassler2007nonstationary}. 
To build the ensemble in our study, we partitioned the data into discrete segments, where each segment serves as an independent realization of the time series. The total length of the time series for the price return is denoted as $N$, while the length of each segment is represented as $S$, so the number of segments is $N_s=N/S$. The chopping autocorrelation is formulated as
\begin{equation}
    \langle C_s \rangle = \frac{C_s^i}{N_s},
\end{equation}
where $C_s^i$ corresponds to the $i$\textsuperscript{th} segment within the ensemble. 
In the typical practice of establishing the ensemble, each segment is often delineated based on individual trading days. However, due to the continuous nature of the Bitcoin market, we chose to employ calendar weeks as partitioning markers for our dataset. Given that our data are recorded at 10-minute intervals, resulting in 6 data points recorded per hour and $1,008$ data points accumulated each week. We rounded this value to 1,000 data points for the length of each segment. Subsequently, we computed the autocorrelation for each segment and averaged it over the ensemble. The error associated with the ensemble autocorrelation was calculated as the standard error within the ensemble.

The results of the ACF calculated with both methods are presented in Figure~\ref{fig:autocor} for Periods 1 and 2, respectively. The sample autocorrelation is plotted in yellow, while the chopping autocorrelation is in blue. 
The absolute sample autocorrelation functions are illustrated in Figure \ref{fig:autocor} (b) and (d) using a log-log scale, corresponding to Periods 1 and 2 respectively. In both time periods, a noticeable power-law relationship emerges, particularly evident at short time intervals (below 100 min). Over the longer term, the absolute autocorrelation exhibits a modest yet nonnegligible value, persisting notably beyond the 200-minute mark ($C(s)=0.002$ for larger $s$). To quantitatively assess this power law behavior, we performed linear fitting, obtaining slope values of $-1.17$ for Period 1 and $-1.07$ for Period 2. 
This slope of absolute autocorrelation is related to the Hurst exponent ($H$), a relationship established as $C(s) \sim s^{-2-2H}$ \cite{Kantelhardt2002multi}. By calculating the Hurst exponent from the power-law slope of the absolute autocorrelation, we obtain the respective values of the Hurst exponent for Periods 1 and 2 as $H=0.415$ and $H=0.486$. Notably, these values align with the findings detailed in Section~\ref{sec:anomalous}.
To further examine the behavior of the short-time autocorrelation, we plotted the ACF for the first 200 minutes in Figure \ref{fig:autocor} (a) and (c). Both sample and chopping autocorrelation show negative values at short time lags, indicating an anti-correlated relationship in short time frames. %Additionally, ACFs manifest periodic oscillations around 0 from 30 minutes to 200 minutes (approximately 2.5 hours), yet the period is different for each time period. 
Additionally, ACFs manifest oscillations around 0 from 30 minutes to 200 minutes (approximately 2.5 hours). Although both the sample autocorrelation and the chopping autocorrelation exhibit similar characteristics in general, they differ in terms of magnitude. Although there are disparities between the two, these differences are not substantial enough to definitively conclude whether they indicate non-ergodic behavior in the time series or are simply a result of statistical noise.
%The discrepancy between the sample autocorrelation and chopping autocorrelation indicates that the Bitcoin price return is non-ergodic. 

%To systematically capture both the power-law and periodic components in the ACF, we employed a two-step calibration approach for the chopping autocorrelation. First, we applied linear regression to determine the slope characterizing the power-law relationship in the absolute chopping autocorrelation. Secondly, we incorporated a cosine function to account for the periodic behavior. The following relation was used to model the ensemble autocorrelation:
%\begin{equation}
%    <\hat{C}(s)>=(cs^a) \text{cos}(bs)
%\end{equation}
%where $b$ and $c$ represent the fitting parameters, and $a$ corresponds to the slope derived from the power-law fitting of the absolute chopping autocorrelation. The least-squared curve fitting was performed to find the parameters, and the resulting fitted curves are plotted in red in Figure \ref{fig:autocor} (a) and (c). Fitting results show that for Period 1, $a=-1.17, b=0.21$, and $c=1.76$, whereas for Period 2, results are $a=-1.07, b=0.13$, and $c=0.40$. Parameter $b$ is related to the periodicity of the observed fluctuation. The fitted results indicate that for Period 1, the period is approximately 60 minutes, whereas in Period 2, the period is 50 minutes.
%===========================================================================
%\section{Stylized facts of detrended stock market time series}\label{sec:detrended}

%===========================================================================
\subsection{Self-similarity of detrended PDF of price return}\label{sec:self_similar}
While the previous section focused on stylized facts within trended time series, it is important to note that the underlying trend inherent in financial data can potentially influence these characteristics. To address this, we now redirect our attention to exploring the stylized facts in the detrended time series. In previous sections, we found that for the Bitcoin market index, the market characteristics for Periods 1 and 2 are similar. From here, we focus the analysis on Period 2 only.

Self-similarity in the evolution of the PDF of price return is another important stylized fact in the stock market. We first tested the self-similarity of the trended time series, yet it does not present clear self-similar behavior. Thus, the detrending process is required following the relation described in Eq \ref{eq:trended_x}. 
\subsubsection{Detrending time series}
The detrending process was carried out using the moving average (MA) method. In MA, a time window is shifted from the start to the end of the time series, and the arithmetic average is used for each time window to record the trend. The two parameters that are vital for MA are the size of the time window $t_w$ and the step in which each time window is shifted forward. In this analysis, we used a continuous sliding window with overlaps, and thus the step is 1. These time windows extend from a segment of $[t, \,\,\, t+t_{w}]$ to the consecutive window of $[t+1, \,\,\, t+t_{w}+1]$.
To achieve an effective detrending result, it is necessary to select an optimal time window $t_w$ for detrending. In this study, the criteria for choosing the optimal time window are set so that the PDFs of the detrended price return $P(x^*,t)$ show the best convergence to a Gaussian distribution at large time intervals $t$ and the goodness of fit ($R^2$) indicates a valid fitting ($R^2\geq0.95$) for all PDFs.

We tested the time window for detrending from 1 hour to 26 weeks to find the best time window that meets the criteria. The optimal time window $t_w$ of 1 week is chosen for the detrending, and for each time window, the arithmetic average of the index $I(t)$ from $[t, \,\,\, t+t_{w}]$ is used as the trend at time $t$. 
Considering at the beginning and the end of the time series where $t<\frac{t_{w}}{2}$ and $t>N-\frac{t_{w}}{2}$ ($N$ is the total length of the time series), the size of the time window is truncated, we define the sliding window with three pieces:

\begin{description}[font=$\bullet$~\normalfont]
\item [For $t< \frac{t_{w}}{2}$] 
\begin{equation}\label{eq:MA1}
\hat{I}(t)=\frac{2}{t_{w}+2t}{\sum_{k=-t+1}^{\lceil(t_{w}-1)/{2}\rceil} I(t+k)},
\end{equation}
\item [For $\frac{t_{w}}{2}<t<N-\frac{t_{w}}{2}$] 
\begin{equation}\label{eq:Ma2}
\hat{I}(t)=\frac{1}{t_{w}}{\sum_{k=-\lfloor(t_{w}-1)/{2}\rfloor}^{\lceil(t_{w}-1)/{2}\rceil} I(t+k)},
\end{equation}
\item [For $t>N-\frac{t_{w}}{2}$] 
\begin{equation}\label{eq:MA3}
\hat{I}(t)=\frac{2}{2N-2t+t_{w}}{\sum_{k=-\lfloor(t_{w}-1)/{2}\rfloor}^{\lceil N-t\rceil} I(t+k)},
\end{equation}
\end{description}
with the time step of $t=1,2,3....N$ for the index fluctuations.

\begin{figure}[!ht]
\centering
\includegraphics[scale=0.6, clip]{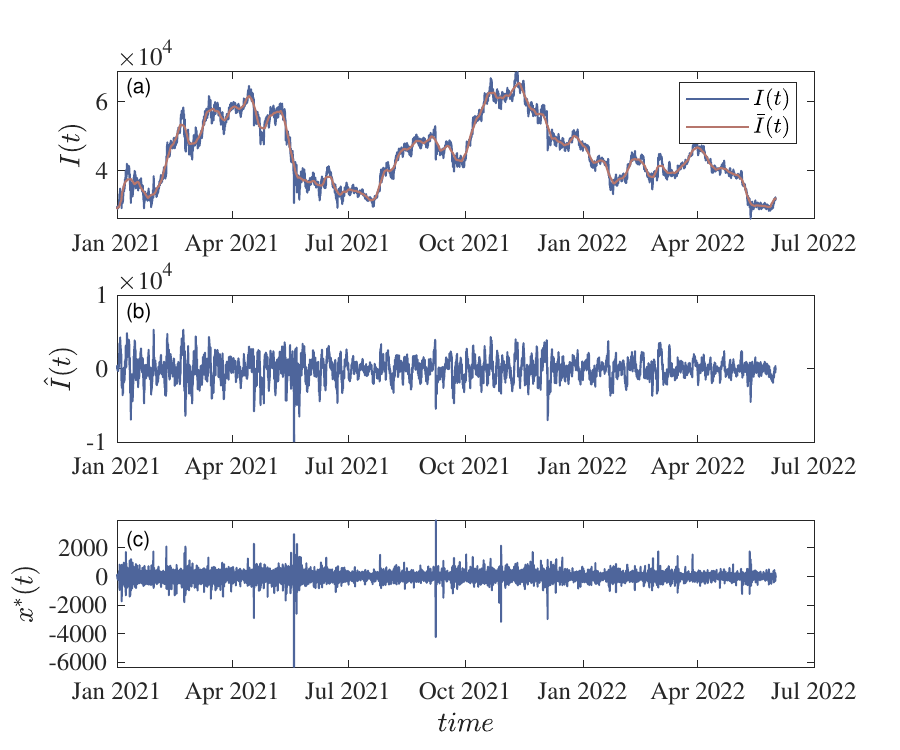}
\caption{Results from detrending analysis for Period 2. (a) Trended price index $I(t)$ is shown as the blue curve, and the trend $\Bar{I}(t)$ obtained from the MA is shown as the red curve. (b) The detrended price index $\hat{I}(t)$ was obtained by subtracting the trend from the price index. (c) Detrended price return $x^*(t)$ calculated from the detrended price index.}
\label{fig:sec3_detrend}
\end{figure}

The results of the detrending are shown in Figure \ref{fig:sec3_detrend}. Figure \ref{fig:sec3_detrend}-a presents the trended market index $I(t)$ as the blue curve and the trend $\Bar{I}(t)$ in red after applying MA analysis using the optimal time window of 1 week. Figure \ref{fig:sec3_detrend}-b shows the detrended price as a result of subtracting the trend, following Eq. \ref{eq:detrend_I}. Figure \ref{fig:sec3_detrend}-c shows the detrended price return by taking the difference of the adjacent terms in the detrended price, using Eq. \ref{eq:increment_price_return}
\begin{align}\nonumber 
    x^*(t) = \hat{I}(t+1) - \hat{I}(t).
\end{align} 

\subsubsection{Self-similarity in PDF of detrended price return}
\begin{figure}[!ht]
\centering
\includegraphics[scale=0.6, clip]{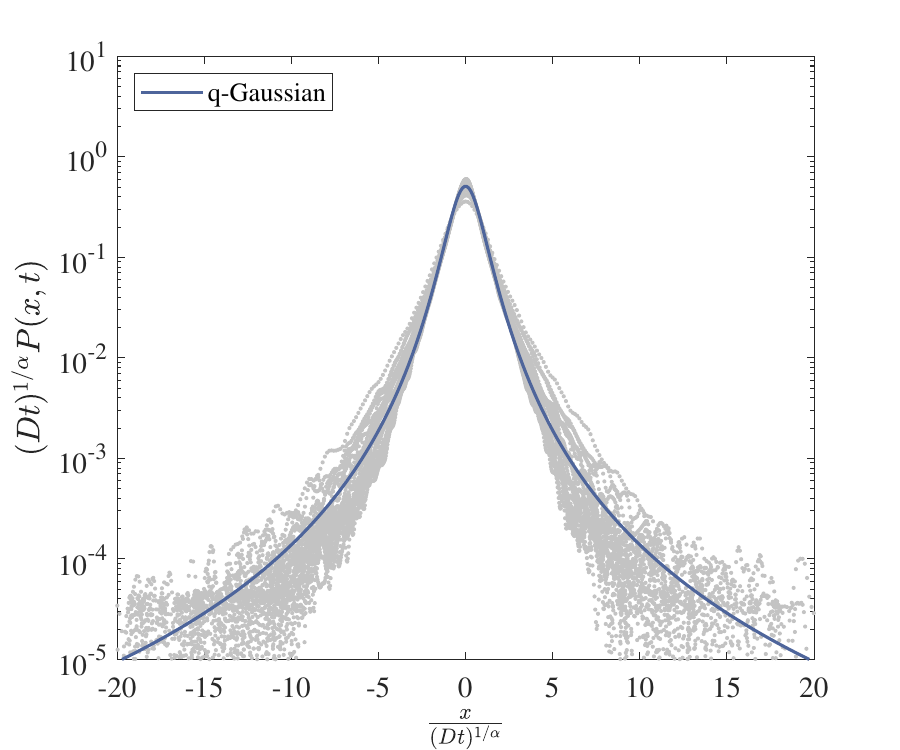}
\caption{(Grey curves) Collapse of the PDF of detrended price return for Bitcoin Period 2. The collapsed PDFs are fitted with a $q$-Gaussian distribution (blue curve) with $q=1.51$.}
\label{fig:self_similar}
\end{figure}

We then test the self-similarity in the detrended price return. Recall the expression of the PDF to be:
\begin{equation}\nonumber
P(x,t)=\dfrac{1}{(D t)^{H}} \left[ g_q \left(\dfrac {x}{(D t)^{H}}\right) \right]
\end{equation}
where $(D t)^{H}$ is the scaling factor, and $H=1/\alpha$ is the Hurst exponent, and is related to the parameter characterizing the anomalous diffusion. Taking the result of $H=0.478$ in Section \ref{sec:anomalous} for time series Period 2, we performed the $q$-Gaussian fitting to the detrended PDFs at each time $t$ of price return on a semi-log scale, with two fitting parameters $q$ and $\beta=(Dt)^{1/\alpha}$ to find the scaling factor. The PDFs are rescaled using these factors, and the resulting PDFs were collapsed onto each other, as shown in the gray curves in Figure \ref{fig:self_similar}. The collapsed PDF shows good agreement with the $q$-Gaussian distribution with $q=1.51$, shown as the blue curve.

%===========================================================================
\subsection{Scaling Analysis on the Fractality of Price Return}\label{sec:DFA}
%A scaling analysis is conducted to find the profile of the Hurst exponent and the fractality of the price return. 
In the preceding sections, we demonstrated the presence of self-similarity in Bitcoin price returns, suggesting the fractal nature of the time series. In this section, our aim is to establish whether the time series is monofractal or multifractal by performing a scaling analysis. In the latter scenario, self-similarity is preserved, but the Hurst exponent is not unique. Instead, it exhibits a range of values that form a Hurst exponent profile. 

Various methods can demonstrate the fractality of a time series, with commonly used approaches including rescaled range analysis (R/S) and detrended fluctuation analysis (DFA). Currently, DFA is becoming a more favored method because of its effectiveness in handling nonstationary time series. In our study, we applied DFA to the Bitcoin Period 2 to determine fractality and calculate the Hurst exponent. DFA was performed on trended time series. The first step of DFA involves removing the trend of the original time series by assuming that the trend is the linear fit of each non-overlapping segment. %In our analysis, this part is only applied to the trended time series, as the trend has already been removed in the detrended time series.
\begin{description}[font=$\bullet$~\normalfont]
\item [\textit{Step 1}] For the trended price return with length $N$, the process of the DFA starts with defining the `profile' of the time series by calculating the mean-centered cumulative sum of the simple price return ($X$):
\begin{equation}
    I^*(t)=\sum_{k=1}^i \left[ X_k -\langle X \rangle \right], i = 1, ... , N.
\end{equation}
where $\langle X \rangle$ is the mean value of the time series.
%For the detrended time series, the profile matches with the detrended index $I^*(t)$. 
\end{description}

\begin{figure*}[htbp]
\centering
\includegraphics[scale=0.35,trim={0 50 0 10}, clip]{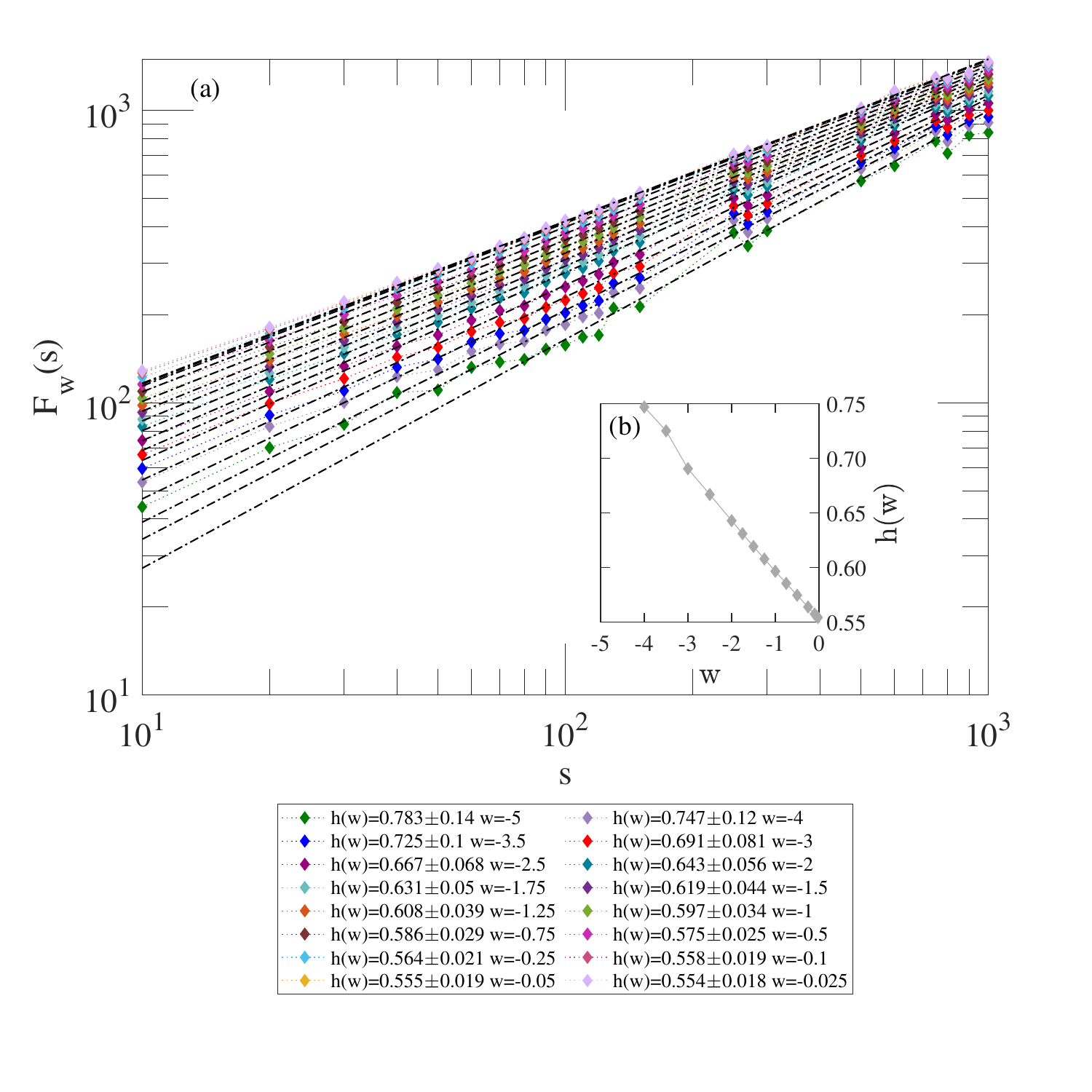}
\includegraphics[scale=0.35, trim={0 50 0 10}, clip]{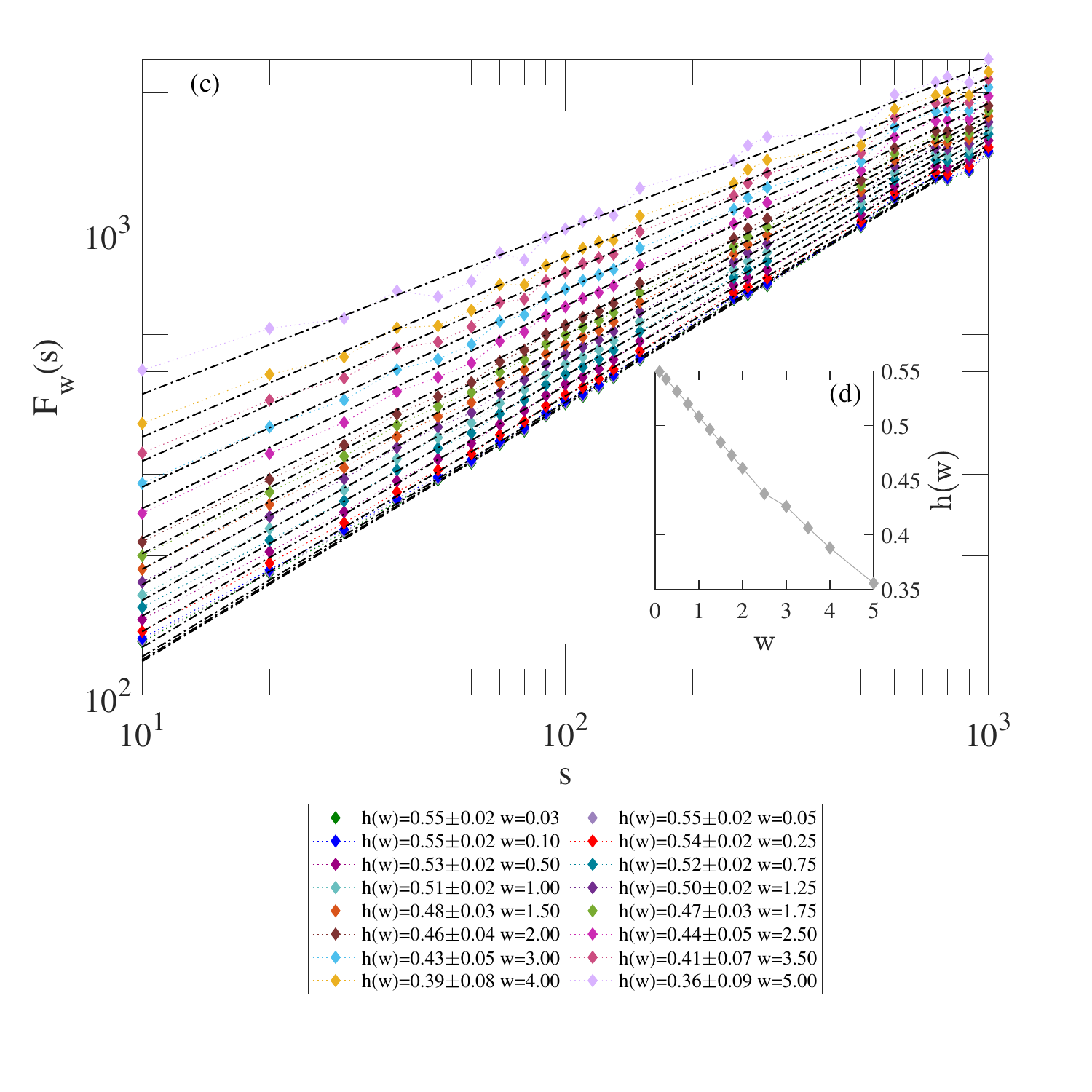}
\caption{Calculation of the statistical function $F_w$ on Bitcoin price return for Period 2 using Eq.~(\ref{eq:DFA3}). The function of $F_w$ vs $s$ display power laws $F_{w}(s) \sim s^{h(w)}$, where $h(w)$ depend on $w$. This feature demonstrates that the time series is multifractal. (a) Calculated $F_{w}(s)$ function with a negative range of $w$, each $w$ value presents a power-law relationship. (b) Profile of $h(w)$ for negative $w$ values. (c) Calculated $F_{w}(s)$ function with a positive range of $w$, each $w$ value presents a power-law relationship. (d) Profile of $h(w)$ for positive $w$ values.}
\label{fig:DFA}     
\end{figure*}

The second part of DFA aims to calculate the scaling function $F_w(s)$ as a function of the time segment $s$, and $w$ is the order of the mathematical moment. This is achieved by applying the following steps:

\begin{description}[font=$\bullet$~\normalfont]
\item [\textit{Step 2}]  Divide the profile $I^*(t)$ into non-overlapping segments with the same length $s$. The number of segments $N_s$ is calculated as $  N_{s}=\lfloor N/s \rfloor $. 

\item [\textit{Step 3}] Calculate the local trend for each segment by a linear least-square fitting of the time series. Then the variance of each segment $v$ from $1,2,3....N_{s}$ is calculated using the equation:
\begin{equation}
F^{2}(v,s)=\frac{1}{s}{\sum_{i=1}^{s} (I^*[(v-1)s+i]-\overline{I^{*}}(v,s))^{2}},
\end{equation}
where $\overline{I^{*}}(v,s)$ is the mean of each segment of $I^{*}(t)$. 

\item [\textit{Step 4}] The statistical moments are calculated utilizing different values of order $w$:
\begin{equation}\label{eq:DFA3}
F_{\mathrm{w}}(s)=\left\lbrace \frac{1}{N_{s}}{\sum_{i=1}^{N{s}} [F^{2}(v,s)]^{\mathrm{w}/2}}\right\rbrace ^{1/\mathrm{w}}
\end{equation}

\end{description}

This moment analysis identifies the spectrum of the time series. For (monofractal or multifractal) scale free time series, the following power law is expected
\begin{equation}
F_w (s) \sim s^{h(w)},
\end{equation}
where the Hurst exponent is $H=h(2)$. For monofractal time series the slope of $h(w)$ in terms of $w$ is constant (independent of $w$), and the fractal dimension of the time series satisfies $D_f=2-H$. Also, the exponent of the autocorrelation function $\gamma$ satisfies $\gamma=2-2H$~\cite{beran2017statistics}. A multifractal time series is characterized by the spectrum of local slopes that depend on $w$. The above functions are related to the general $w$th moment defined as follows
\begin{equation}
G_w(s)\equiv {\sum_{v=1}^{N/s} |I^*[vs]-I^*[(v-1)s]|^{w}}
\end{equation}
which defines the generalized scaling exponent $\tau(w)$ via the scaling relation
\begin{equation}
G_w(s)\sim s^{\tau(w)}.
\end{equation}
$\tau(w)$ is shown to be related to $h(w)$ via the relation
\begin{equation}
    \tau (w)=wh(w)-1.
    \label{eq:tau}
\end{equation}
\begin{figure}[ht]
\centering
\includegraphics[scale=0.6, clip]{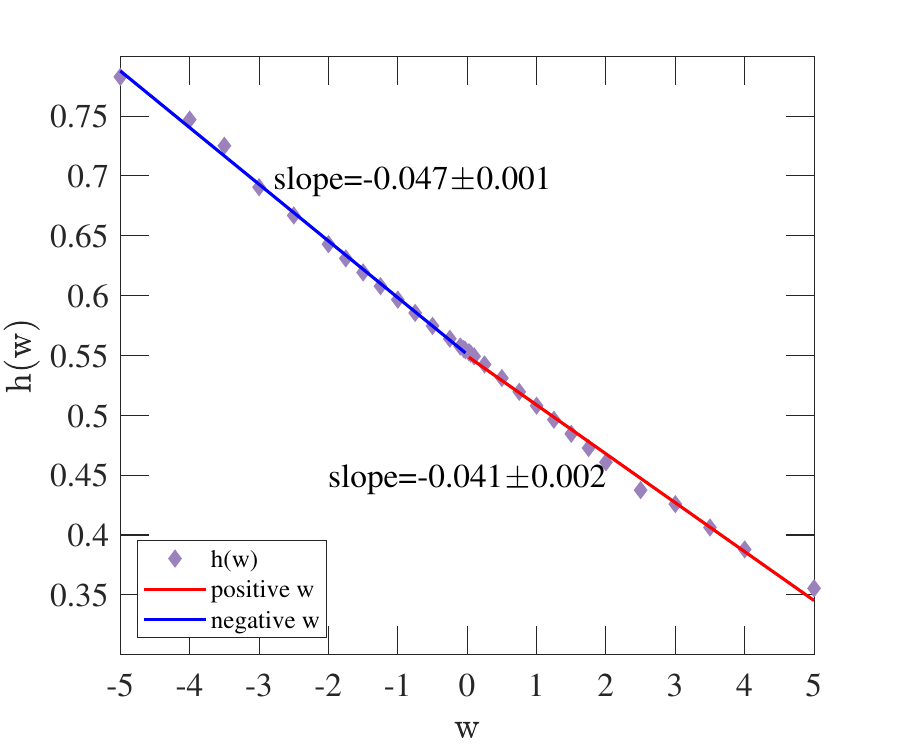}
\caption{Evaluation of the scale exponents $h(w)$ of detrended Bitcoin price return for Period 2. This profile of scale exponent can be described using linear relationships for positive and negative $w$, with the slope being $-0.047 \pm 0.001$ for negative $w$, and $-0.041 \pm 0.002$ for positive $w$, respectively.} %The Hurst exponent is equivalent to the $h(2)$ value, from which $H=0.461$. 
\label{fig:hw}     
\end{figure}

\begin{figure}[ht]
\centering
\includegraphics[scale=0.6, clip]{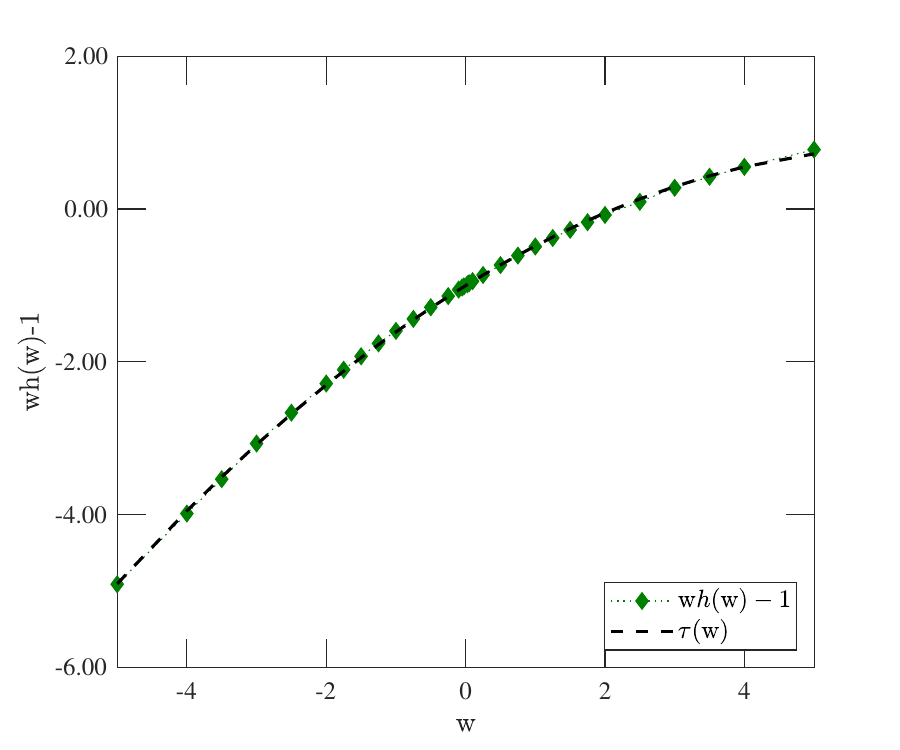}
\caption{Generalized scaling exponent $\tau
(w)$ of the detrended price return for Period 2, showing the relation $\tau (w)=wh(w)-1$.} %The Hurst exponent is equivalent to the $h(2)$ value, from which $H=0.461$. 
\label{fig:tauw}     
\end{figure}

The DFA results are shown in Figure \ref{fig:DFA} for the positive and negative range of $w$, showing a power law behavior. The exponents of the scaling behaviors are extracted by using power-law fitting (dashed black lines), which is carried out for each order of $w$. The insets show $h(w)$ in terms of $w$. It is clear that the slope of $h(w)$ varies for different $w$, indicating that the time series is multifractal. Figure \ref{fig:hw} shows more clearly the resultant generalized Hurst exponent $h(w)$. Figure \ref{fig:tauw} presents the relationship between $\tau (w)$ in terms of $h(w)$, provided that the detrended price return is self-similar and multifractal, since the slope depends on the scale $w$. \\

\begin{figure}[ht]
\centering
\includegraphics[scale=0.55, clip]{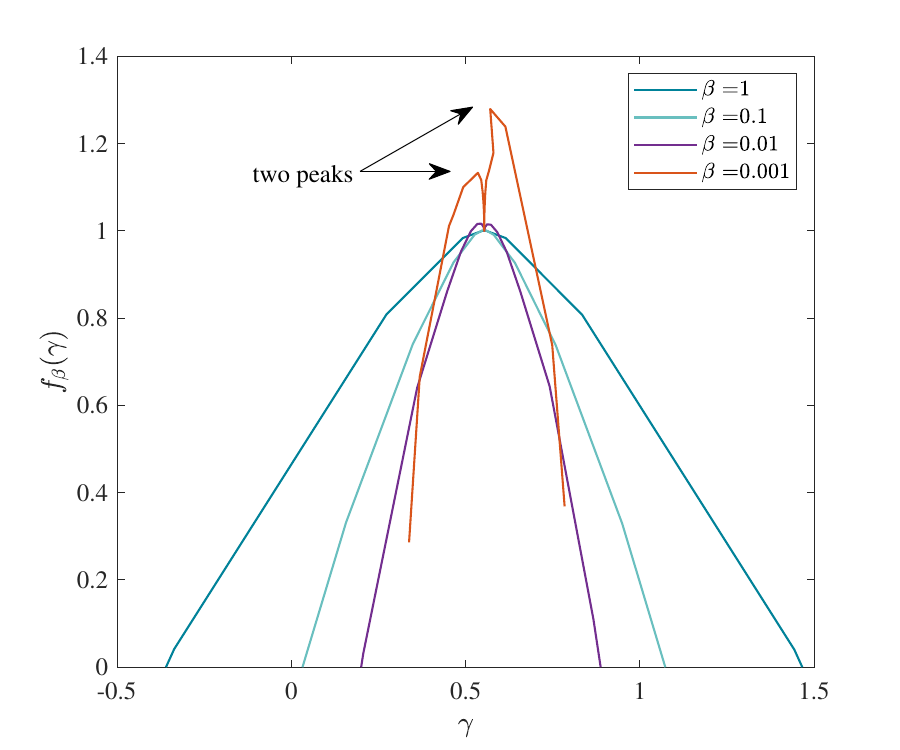}
\caption{Result for $f_{\beta}(\gamma)$ with varying $\beta$ values from 1 to 0.001. As $\beta$ decreases, $f_{\beta}(\gamma) $ becomes narrower. When $\beta =0.001$, two peaks are presented as shown in the red curve.  } %The Hurst exponent is equivalent to the $h(2)$ value, from which $H=0.461$. 
\label{fig:f_gama}     
\end{figure}

To obtain the spectrum of the Hurst exponent, we have to use a Legendre transformation after performing the detrended fluctuation analysis. In the standard theory of multifractal analysis, one performs the Legendre transformation of the generalized scaling exponent $\tau (w)$ as a function of $h(w)$, which is based on the linear relation Eq.~\ref{eq:tau}. The Legendre transformation of $\tau (w)$ is then given by
\begin{equation}
    f(\gamma) = \gamma w-\tau (w)
    \label{eq:f_gama}
\end{equation}
where $\gamma$ is defined as
\begin{equation}
    \gamma=\frac{\partial \tau (w)}{\partial w} = h(w)- w \frac{\partial h}{\partial w}.
    \label{eq:gamma2}
\end{equation}
By substituting Eq. \ref{eq:tau} in Eq. \ref{eq:f_gama}, the Legendre transformation of $\tau (w)$ is found to be
\begin{equation}
    f(\gamma)=w(\gamma -h(w)) +1.
    \label{Eq:Legendre}
\end{equation}
For the case $h(w)$ is linear with respect to $w$, $\gamma$ is found to be constant, making the Legendre transformation ill-defined. This indicates that $\gamma$ is not a one-to-one function of $w$, therefore, not invertible to calculate $f(\gamma)$. To avoid this problem, we add a non-linear auxiliary term to the function of $h(w)$. More precisely, one adds a nonlinear term to $h(w)$, performs the calculations, and sends the amplitude of the linear term to zero at the end. A similar scenario applies here since the deviations from linearity for $h(w)$ are small, i.e. the Legendre transformation results in a constant $\gamma$ with some fluctuations around. To proceed, we add the following non-linear term
\begin{equation}
h_{\beta}(w)=\beta w^3+h(w),
\end{equation}
where $\beta$ is an amplitude. We considered $w^3$ to make the Legendre transformation one-to-one, that is, $f_{\beta}(\gamma)$ (obtained using Eq.~\ref{Eq:Legendre} with $h(w)$ replaced by $h_{\beta}(w)$) is a one-to-one function of $\gamma$. This makes Eq.~\ref{eq:gamma2} invertible, so that $\gamma$ is easily obtained as a function of $w$ to be used in the other equations. In the end, we should take the limit $\beta\to 0$, i.e. 
\begin{equation}
f(\gamma)=\lim_{\beta\to 0}f_{\beta}(\gamma).
\end{equation}
As a simple example to see how this scenario works, consider
the simple linear case $h(w)=aw+b$, so that $h_{\beta}(w)=\beta w^3+ax+b$, which gives
\begin{equation}
    \gamma= -2 \beta w^3+b.
\end{equation}
or $w=(\frac{b-\gamma}{2\beta})^{1/3}$, and $f_{\beta}(\gamma)$ is calculated as:
\begin{equation}
\begin{split}
    &f_{\beta}(\gamma) =w(\gamma -h_{\beta}(w)) +1 \\
    &= \left( \frac{b-\gamma}{2\beta} \right)^{1/2} \left( \gamma-\beta \left( \frac{b-\gamma}{2\beta} \right) -a \left( \frac{b-\gamma}{2\beta} \right) -b \right) +1
\end{split}
\end{equation}
This relation allows us to track and monitor how $f(\gamma)$ approaches $f_{\beta}(\gamma)$ as $\beta$ goes to zero. Figure~\ref{fig:f_gama} shows the result of $f_{\beta}(\gamma)$ with different values of $\beta$. The results show that as $\beta$ approaches the limit of 0, two peaks appear, which are shown in the figure. The two peaks represent the two-fractal behavior of the time series. As we already observed in Fig.~\ref{fig:hw}, the slopes for positive and negative values of $w$ are slightly different, so one may expect two different classes of exponents, which is consistent with the result that we found based on the nonlinear analysis of Fig.~\ref{fig:f_gama}. Therefore, we conclude that the time series studied here is multifractal.

%, where $h(2)$ is the Hurst exponent representing the time series \cite{Kantelhardt2002multi}. For Period 2, $H=0.461$, which agrees with the results obtained in the previous sections. 

%===========================================================================
\newpage
\section{Discussion}\label{sec:Discussion}
\begin{table*}[hbt!]
\centering
\caption{Summary of stylized facts}
\label{tab:summary}
\begin{tabular}{|l|l|l|l|}
\hline
          Stylized Facts              & BTC Period 1 & BTC Period 2 & S\&P500 \footnote{The values presented for the S\&P500 correspond to the analyzed period from 1996 to 2018 per minute\cite{alonso2019q,Arias-Calluari2022stationarity}.} \\ \hline
\begin{tabular}[c]{@{}l@{}}Fat-tailed Distribution\\ $P(x,t) \sim x^{\frac{2}{1-q}}$ for large $x$ \end{tabular}  & $q=1.51$       & $q=1.50$     &  $q\approx1.70$     \\ \hline
\begin{tabular}[c]{@{}l@{}}Anomalous Diffusion\\ (Short time-intervals)\\ $P(x=0, t) \sim t^{-H}, H=1/\alpha $\end{tabular}     & \begin{tabular}[c]{@{}l@{}} $H=0.415$ \\  $\alpha=2.41$ \end{tabular}         & \begin{tabular}[c]{@{}l@{}}  $H=0.478 $\\$\alpha=2.09$ \end{tabular}      & \begin{tabular}[c]{@{}l@{}}  $H=0.79 $ \\$\alpha=1.26$\end{tabular}          \\ \hline
\begin{tabular}[c]{@{}l@{}}$q$-Gaussian Diffusion \\ $P(x,t)=\dfrac{1}{(D t)^{H}} \left[ g_q \left(\dfrac {x}{(D t)^{H}}\right) \right]$\end{tabular} &$q=1.53$        & $q=1.57$   & \begin{tabular}[c]{@{}l@{}}  $q=1.71$\\$q=2.73$ \end{tabular}           \\ \hline
 \begin{tabular}[c]{@{}l@{}}Volatility Clustering\\ $C(s) \sim (cs)^{2H-2} \text{cos}(bs) $\end{tabular}   & $H=0.415$   & $H=0.465$ & \begin{tabular}[c]{@{}l@{}}   $C(s)\sim e^{-\rho s}$ \\$\rho=1.07$ \end{tabular}   \\ \hline
\begin{tabular}[c]{@{}l@{}}Scaling Analysis \\ $F_{w}(s) \sim s^{h(w)}$ \end{tabular} & / & $H=0.461$  & $H=0.48$  \\ \hline
 
\end{tabular}

\end{table*}
We investigated the stylized facts of the Bitcoin time series from 2019 to 2022. The Bitcoin price index was divided into two periods based on changes in volatility. Table~\ref{tab:summary} summarizes the stylized facts observed in the Bitcoin price index. Here we compare the results of the Bitcoin price index with the well-studied conventional market S\&P500 for the period January 1996 to May 2018.

The prevalence of a heavy-tailed distribution is a characteristic feature in traditional financial markets, such as the S\&P500. In line with this, we found fat-tailed distributions for both time periods of Bitcoin price return. This finding aligns with previous studies on cryptocurrencies, where most types of cryptocurrencies exhibit non-Gaussian heavy-tailed distributions \cite{Zhang2018stylized}. Our study extends previous findings by illustrating that the PDF of Bitcoin's price return is best captured by a $q$-Gaussian distribution, making these results directly comparable with our previous results for the S\&P500. Specifically, we find $q=1.51 \pm 0.02$ for Period 1 and $q=1.50 \pm 0.02$ for Period 2.

In comparison to the S\&P500 index, the anomalous diffusion of Bitcoin presents distinctive characteristics. Although Bitcoin and the S\&P500 have undergone a change in diffusion behavior, their respective diffusion patterns differ. For the S\&P500 price return, research has shown that it shows a strong superdiffusion regime with $\alpha=1.26\pm0.04 \, (H=0.79)$ for short time intervals, and in the long term it exhibits a weak superdiffusion regime with $\alpha=1.79\pm0.01 \, (H=0.56)$ \cite{alonso2019q}. In contrast, the Bitcoin price returns exhibit a transition in diffusion behavior in both periods, shifting from a subdiffusion regime to a weak superdiffusion regime. 
In Period 1, $\alpha=2.41\pm0.02 \, (H=0.415)$ for short time intervals and $\alpha=1.64\pm0.02 \, (H=0.610)$ for extended time intervals. In Period 2, short time intervals reveal $\alpha=2.09\pm0.01 , (H=0.478)$, while extended time intervals exhibit $\alpha=1.54\pm0.01 , (H=0.646)$. This suggests that Bitcoin exhibits antipersistent behavior over short time intervals, resisting the formation of trends. However, on longer time scales (up to 5 hours), the Hurst exponent exceeds 0.5, indicating the presence of trends. In particular, the diffusion parameters indicate that Period 1 of Bitcoin demonstrates a more pronounced subdiffusion than Period 2 for short time intervals. The anomalous diffusion of Bitcoin can be described by a $q$-Gaussian diffusion process, similar to the findings of S\&P500. However, the values of the parameter $q$ are different between the two markets. For S\&P500, the $q$-Gaussian exponents are $q=1.71 \pm 0.01$ and $q=2.73 \pm 0.005$ for the two diffusion regimes, respectively \cite{alonso2019q}.  Conversely, for Bitcoin, the $q$-Gaussian exponents are notably lower, at approximately $q=1.55 \pm 0.02$ for Periods 1 and 2. Indicating than extreme returns (both gains and losses) happen more frequently in S\&P500 than in Bitcoin. 

%Our results present a transition in diffusion mode in the Bitcoin market index. This transition is also observed in a more recent study on S\&P500 price index. However, the diffusion mode is different between the two markets. Over short time intervals, S\&P500 price index exhibits a strong superdiffusion regime with $\alpha=1.26\pm0.04 \, (H=0.79)$, while in the long term, it demonstrates a weak superdiffusion regime with $\alpha=1.79\pm0.01 \, (H=0.56)$ \cite{alonso2019q}. For Bitcoin, the diffusion mode changes from a subdiffusion regime to a weak superdiffusion regime, and the diffusion follows a $q$-Gaussian diffusion mode.
The autocorrelation functions are calculated on the basis of the price returns of Bitcoin. Regarding the ACF for absolute returns, it demonstrates a power-law decay, with the calculated Hurst exponents shifting from less than 0.5 in Period 1 ($H=0.415$) to slightly closer to 0.5 in Period 2 ($H=0.486$). In terms of the ACF of price returns, both time periods exhibit short-term negative autocorrelation. %This negative autocorrelation rapidly diminishes and fluctuates around zero, with different amplitudes and periods for each time period. 
Autocorrelation in small intraday time scales (less than $20$ minutes) can be attributed to the microstructure effects of the market. This negative autocorrelation aligns with the findings on bitcoin feedback traders \cite{Karaa2021feedback}, confirming that the high volatility and positive trading strategy can produce significant anticorrelations in market dynamics. The ACF pattern observed in the Bitcoin price return differs from the observations in S\&P500; in the case of S\&P500, the ACF presents an exponential pattern at short times and then transitions to a power-law relationship \cite{Arias-Calluari2022stationarity}. 

%The negative short-time anti-autocorrelation presented in Bitcoin can be explained by the possible trader activities and the high volatility in the time series. Previous studies proposed the connection between the presence of autocorrelation and trader behaviors by devising a model with two types of traders: positive and negative. Positive traders will result in negative autocorrelations, and the converse holds for negative traders. Empirical evidence supporting the argument of how feedback traders can affect autocorrelation is observed across various stock markets \cite{Sentana1992, Koutmos1997, Booth1998volatility}. High volatility in this time period can also lead to negative autocorrelations since a negative relationship is presented between volatility and autocorrelation, also linked by trader behaviors. Studies found that an increase in volatility tends to increase the degree of positive feedback trading, which in turn induces negative return autocorrelation \cite{Black1988Equilibrium, Black1990Mean, McKenzie2003Determinants, McKenzie2007evidence, faff2007volatility}. \Red{Note: The clarity of the previous paragraph can be enhanced by improving Section II. Instead of citing references in the conclusions, these should be used to understand the internal mechanisms of Bitcoin.}

From the scaling analysis, we found that both time periods of the Bitcoin price index are multifractal, which is in agreement with other studies on the fractality of the Bitcoin market \cite{Takaishi2018multifractality, Shrestha2021Multifractal}. This multifractality characteristic is similar to the observations in traditional financial markets and other complex systems \cite{Lee2017Asymmetric, Mali2014Multifractal, Wang2013Multifractal, Gu2010Multifractal, Chen2016Finite-size}. The multifractal nature of Bitcoin can be attributed to the presence of volatility clusters varying across different time scales and fat-tailed distributions, as these aspects make essential contributions to multifractality in a given time series \cite{Lahmiri2018Chaos}.

%% The Appendices part is started with the command \appendix;
%% appendix sections are then done as normal sections
%% \appendix

%% \section{}
%% \label{}

%% References
%%
%% Following citation commands can be used in the body text:
%% Usage of \cite is as follows:
%%   \cite{key}         ==>>  [#]
%%   \cite[chap. 2]{key} ==>> [#, chap. 2]
%%

%% References with BibTeX database:

\bibliographystyle{ieeetr}

%\bibliography{references}
%% Authors are advised to use a BibTeX database file for their reference list.
%% The provided style file elsarticle-num.bst formats references in the required Procedia style

%% For references without a BibTeX database:

% \begin{thebibliography}{00}

%% \bibitem must have the following form:
%%   \bibitem{key}...
%%

% \bibitem{}

% \end{thebibliography}

\end{document}